# Report of the Medical Image De-Identification (MIDI) Task Group - Best Practices and Recommendations

David A. Clunie, Adam Flanders, Adam Taylor, Brad Erickson, Brian Bialecki, David Brundage, David Gutman, Fred Prior, J Anthony Seibert, John Perry, Judy Wawira Gichoya, Justin Kirby, Katherine Andriole, Luke Geneslaw, Steve Moore, TJ Fitzgerald, Wyatt Tellis, Ying Xiao, Keyvan Farahani

## 1.1 Task Group

| Name | Affiliation | ORCID |
|---|---|---|
| Adam Flanders | Thomas Jefferson University | 0000-0002-4679-0787 |
| Adam Taylor | Sage Bionetworks | 0000-0003-0501-8886 |
| Brad Erickson | Mayo Clinic | 0000-0001-7926-6095 |
| Brian Bialecki | American College of Radiology | 0000-0003-1835-7449 |
| David Brundage | Cornell University | |
| David Clunie | PixelMed Publishing | 0000-0002-2406-1145 |
| David Gutman | Emory University | 0000-0002-1386-8701 |
| Fred Prior | University of Arkansas for Medical Sciences | 0000-0002-6314-5683 |
| J Anthony Seibert | University of California, Davis | 0000-0002-0606-0990 |
| John Perry | Independent consultant | 0000-0001-5833-2065 |
| Judy Wawira Gichoya | Emory University | 0000-0002-1097-316X |
| Justin Kirby | Frederick National Laboratory for Cancer Research | 0000-0003-3487-8922 |
| Katherine Andriole | Brigham and Women's Hospital | 0000-0001-7774-4225 |
| Luke Geneslaw | Memorial Sloan Kettering Cancer Center | 0000-0001-6862-9102 |
| Steve Moore | Washington University in St. Louis | 0000-0002-1952-3114 |
| TJ Fitzgerald | UMass Memorial Medical Center | 0000-0002-6256-0615 |
| Wyatt Tellis | University of California, San Francisco | 0000-0001-8818-2379 |
| Ying Xiao | University of Pennsylvania Health System | 0000-0001-8558-6394 |
| Keyvan Farahani | National Institutes of Health | 0000-0003-2111-1896 |

## 1.2 Advisory Group

| Name | Affiliation | ORCID |
|---|---|---|
| James Luo | National Heart, Lung, and Blood Institute (NHLBI) | |
| Alex Rosenthal | National Institute of Allergy and Infectious Diseases (NIAID) | 0000-0003-4190-9045 |
| Kris Kandarpa | National Institute of Biomedical Imaging and Bioengineering (NIBIB) | |

| Rebecca Rosen | Eunice Kennedy Shriver National Institute of Child Health and Human Development (NICHD) | 0000-0002-0680-6246 |
|---|---|---|
| Kerry Goetz | National Eye Institute (NEI) | 0000-0002-9821-7704 |
| Debra Babcock | National Institute of Neurological Disorders and Stroke (NINDS) | |
| Ben Xu | National Institute on Alcohol Abuse and Alcoholism (NIAAA) | |
| John Hsiao | National Institute on Aging (NIA) | 0000-0003-3471-7539 |

## 1.3 Acknowledgements

Leadership: Keyvan Farahani ORCID 0000-0003-2111-1896, Center for Biomedical Informatics & Information Technology (CBIIT), NCI
Project management: Ulli Wagner, Frederick National Laboratory for Cancer Research
Technical writer: Carolyn Kelley Klinger, Frederick National Laboratory for Cancer Research
Advice on SDC: Khaled El Emam, University of Ontario
External review: Benjamin Kopchick, Deloitte Consulting
External review: Peter Kuzmak, Veterans Health Administration
External review: Michael Onken
External review: Carlos Costa ORCID 0000-0002-2707-5331, Luís Bastião ORCID 0000-0001-8513-7185, Rui Jesus ORCID 0000-0002-9231-6744 and Jorge Silva ORCID 0000-0002-6331-6091
External review: Liam Caffery, University of Queensland ORCID 0000-0003-1899-7534
External review: William Parker, University of British Columbia
External review: Liron Pantanowitz, University of Michigan ORCID 0000-0001-8182-5503
External review: Peter van Ooijen, University Medical Center Groningen ORCID 0000-0002-8995-1210
External review: Markus Daniel Herrmann, Massachusetts General Hospital ORCID 0000-0002-7257-9205
External review: Andrey Fedorov, Brigham and Women's Hospital ORCID 0000-0003-4806-9413
External review: Toine van Wijk, ICT Netherlands
External review: Christopher Schwarz, Mayo Clinic ORCID 0000-0002-1466-8357
External review: For ADNI, John Hsiao ORCID 0000-0003-3471-7539, Christopher Schwarz ORCID 0000-0002-1466-8357, Arthur Toga ORCID 0000-0001-7902-3755, Paul Yushkevich ORCID 0000-0001-8543-4016, Duygu Tosun-Turgut ORCID 0000-0001-8644-7724, Michael Weiner ORCID 0000-0002-0144-1954
External review: Petr Vcelak ORCID 0000-0003-4415-790X
External review: Paul Kinahan ORCID 0000-0001-6461-3306

## 1.4 Support

This project has been funded in whole or in part with Federal funds from the National Cancer Institute, National Institutes of Health, under Contract No. 75N91019D00024, Task Order 75N91019F00129. The content of this publication does not necessarily reflect the views or policies of the Department of Health and Human Services, nor does mention of trade names, commercial products, or organizations imply endorsement by the U.S. Government.

## 1.5 Summary of Scope

This report addresses the technical aspects of de-identification of medical images of human subjects and biospecimens, such that re-identification risk of ethical, moral, and legal concern is sufficiently reduced to allow unrestricted public sharing for any purpose, regardless of the jurisdiction of the source and distribution sites.

All medical images, regardless of the mode of acquisition, are considered, though the primary emphasis is on those with accompanying data elements, especially those encoded in formats in which the data elements are embedded, particularly Digital Imaging and Communications in Medicine (DICOM). These images include image-like objects such as Segmentations, Parametric Maps, and Radiotherapy (RT) Dose objects.

The scope also includes related non-image objects, such as RT Structure Sets, Plans and Dose Volume Histograms, Structured Reports, and Presentation States.

Except to the extent that the handling of standard DICOM information model attributes such as Patient, Study, Series, and Instance identifiers and descriptors is similar to that of images, the following are not in scope, in the sense that this report does not specifically consider their specific aspects:

- Raw data, including MR k-space, CT projections, and PET list mode data
- Encapsulated payloads, such as CDA and PDF documents, and 3D models in STL or OBJ.
- Surface segmentations
- Physiological time-based waveforms, including ECG, EEG, and others
- Audio waveforms

Only de-identification of publicly released data is considered, and alternative approaches to privacy preservation, such as federated learning for artificial intelligence (AI) model development, are out of scope, as are issues of privacy leakage from AI model sharing.

Only technical issues of public sharing are addressed. Other matters such as the following are out of scope:

- policies,
- controls, such as data use agreements,
- registered and limited access restrictions,
- pseudonymization procedures,

- retention of identity maps at the submitting site or elsewhere,
- encryption in transit and at rest during the de-identification process,
- mitigation of privacy breaches

## 1.6 Summary of Best Practices

1. Thorough de-identification by removal or replacement of all known direct and relevant indirect identifiers and sensitive information, in all collection descriptions and supporting data, structured and unstructured text data elements, pixel data, and geometric and bitmapped overlays, is required for public sharing. Direct identifiers should always be removed. A realistic collection-specific expert statistical analysis should be performed to quantify residual re-identification risk with respect to a pre-determined risk threshold, to justify retention of selected indirect identifiers or sensitive information, potentially with modified risk-reduced values, to preserve re-use utility. Any such risk analysis needs to consider any other publicly available information about the subject, and is only valid at the point in time at which it was done; consideration should be given to the potential for an increase in risk over time.
2. To maximize retention of scientific value, re-use utility should be defined with respect to the broadest range of possible use cases conceivable, including those beyond the scope of the entity releasing the data; quantification of utility should be performed if possible when statistically assessing residual re-identification risk.
3. The de-identification process should not compromise the conformance of the resulting data with the standards that define the content, or reduce the level of functionality; specifically, de-identification of DICOM files should retain DICOM conformance with the original information object definition (IOD), even if that requires synthesis of dummy values for replacement, and consistent replacement values across multiple files (e.g., to retain referential integrity of replaced unique identifiers within a defined scope). This requires retention or replacement of not only required attributes, but also optional attributes critical to retain functionality.
4. The de-identification process should preserve as much information about the image acquisition as possible (including machine identity, characteristics, and settings) to maximize the re-use potential, except to the extent that machine information can be realistically quantified as increasing the residual re-identification risk above a pre-determined acceptable risk threshold.
5. Though patient re-identification risk is the primary concern, the identity of other persons, departments and institutions, and organizations may pose privacy concerns and/or be a source of indirect identifiers or sensitive information about the patient. This information should be removed or replaced except as necessary to preserve re-use utility, subject to a residual re-identification risk assessment.
6. For DICOM images, the current release (at the time of de-identification) of the DICOM PS3.15 E.1 Application Level Confidentiality Profile should be used as a reference for those structured and unstructured data elements that need to be de-identified, augmented by any additional knowledge of other unsafe attributes, including private data elements, that need to be considered. Data elements that are retained to preserve utility, whether standard or private, should be categorized per the DICOM PS3.15 options. The PS3.15 approach of removing or replacing everything that is known to be unsafe, and retaining only what is known to be safe, is not inherently information-object-specific, in that it is applicable to any DICOM object, whether an image or not. The specific DICOM IOD does not need to be "recognized" to be processed in this manner (e.g., a Raw Data object processed in this manner would have all its patient identifiers replaced, and its private raw data elements containing the raw data payload removed unless the latter were definitively known to be safe). DICOM PS3.15 defines various options beyond the baseline for retention, cleaning, or removal of information for various scenarios, and these choices should be carefully evaluated to balance preservation of utility against residual re-identification risk. E.g., text descriptions at the study or series level may be useful to select appropriate images but may leak identity and hence need to be removed unless specific cleaning processes, whether manual or automated, are implemented. E.g., patient characteristics may need to be retained to make PET images useful for quantification, but these potentially create re-identification risk, being indirect identifiers.

7. If any patient characteristic or other indirect identifier data elements are retained, a collection-specific expert statistical analysis of residual re-identification risk should be performed. This applies regardless of whether data elements are selectively retained (allowlist, whitelist) or selectively removed (denylist, blacklist). No arbitrary finite number or specified subset of indirect identifiers is considered acceptable to retain in the absence of a risk analysis. The possibility of deriving patient characteristics (such as sex or age) from the pixel data of the images themselves should be considered.
8. For non-DICOM images, in the absence of an alternative specific reliable reference for data element retention or removal, the general principles explicit or implicit in DICOM PS3.15 E.1 should be applied, e.g., for images stored in DICOM-derived formats like Brain Imaging Data Structure (BIDS) with an alternative metadata representation. For clinical data elements, the general principles in the PhUSE De-Identification Standard for CDISC SDTM should be applied.
9. Regardless of the image encoding or file format, all data elements linked to images in the collection, including those in accompanying spreadsheets or publications, which are linked by a common key (e.g., the pseudonymous subject identifier) need to be de-identified and subject to a risk analysis. That risk analysis should account for linked information in other public data sets for the same subjects, which are made available by other organizations and that are known to the de-identifier. E.g., Genomic Data Commons (GDC) may make indirect identifiers or sensitive information available for The Cancer Genome Atlas (TCGA) subjects, even if that information is not present or has been removed from images and accompanying spreadsheets shared by The Cancer Imaging Archive (TCIA). A search for the existence of such linked data should be undertaken.
10. The risk posed by the presence of burned-in text, foreign objects with textual information (e.g., jewelry) and other sources of potential identity leakage (e.g., face, fingerprint, tattoo) in pixel data should be assessed, and if the risk exceeds a pre-determined threshold, scanned for the offending information, and the entire image discarded or the offending information redacted, manually or automatically (subject to subsequent human review); the effort to scan and redact versus discard should be weighed against re-use utility. This risk assessment should be performed for all image types, no matter what the mode of acquisition, including but not limited to radiographic cross-sectional and projection, ultrasound, nuclear medicine and visible light (including external photographic, microscopic and endoscopic), screen and video capture, any of which may contain physical elements with identification, including labels attached to the patient as well as digitally burned-in identifying annotations. It is not sufficient to limit checks for offending information to only a stratified sub-set of image types (such as might be determined from other data elements, including modality, explicit type information and various descriptive fields), though some may be of higher risk than others.

11. All non-structured text fields, whether in data elements or burned-in to pixel data or graphical or bitmapped overlays, should be either completely removed or replaced, or if needed for re-use utility, scanned for embedded direct and indirect identifiers and sensitive information, either manually or automatically (subject to subsequent human review).
12. Absolute dates and times should be replaced, but relative temporal integrity across imaging studies performed at different time points should be preserved, within a useful tolerance, in order to retain utility on a short time frame for dynamic contrast or nuclide decay computation, and on a longer term for therapy assessment or other clinically relevant interval change.
13. Private data elements retained to preserve utility should be evaluated with respect to risk of identity leakage, either by reference to a reliable source of known safe private data elements, such as that provided in DICOM PS3.15 E.3.10, manufacturer's documentation, including DICOM Conformance Statements, or published documents from other reliable sources. Otherwise, private data elements should be selectively or entirely removed.
14. Compressed bitstreams used as pixel data or within other data elements (e.g., International Color Consortium (ICC Profiles)) should be considered with respect to the potential for identity leakage through embedded data elements, and either decompressed during de-identification (if losslessly compressed) and the embedded data elements discarded, or if the compressed bitstream is re-used, scanned for data elements at risk and those selectively removed or replaced. E.g., an EXIF APP1 or JUMBF APP11 marker segment in the lossy JPEG pixel data of a DICOM image may contain direct or indirect identifiers in data elements as well as information of re-use utility.
15. The re-identification risk of head and neck cross-sectional images, including brain CT, MR and PET images, which may contain potentially reconstructable facial information (PRFI) that can be used by humans or facial recognition software to attempt re-identification, should be quantified with a realistic collection-specific expert statistical analysis, and if above a predetermined acceptable risk threshold, the facial features removed or modified to reduce the risk to an acceptable level, or the images should not be publicly shared.

16. Photographic images that include faces should be redacted in such a manner as to confound attempts by humans or facial recognition software at re-identification, quantified with a realistic collection-specific expert statistical analysis, and if above a predetermined acceptable risk threshold not publicly shared.
17. A human quality control (QC) process to confirm the efficacy of the de-identification process used with respect to de-identification and preservation of utility should be used; the percentage and type of records inspected should be guided by a documented risk assessment establishing the threshold of residual risk before and after performance of the QC process. The QC process should address structured and unstructured text data elements, pixel data, geometric and bitmapped overlays, and compressed bitstream embedded metadata. The residual risk is influenced by the assessment of what is to be removed or replaced, as well as the reliability of the manner in which it is removed or replaced.
18. The process of de-identification used, including that performed by source sites, data coordinating centers and the entity that is responsible for the public data distribution, should be documented in detail, and that documentation, or a reference to an openly accessible source of it, published with the data collection. This documentation should include the release of the PS3.15 E.1 Application Level Confidentiality Profile used, as well as documenting any PS3.15 Confidentiality Options used.

## 1.7 Summary of Recommendations

1. Continued maintenance of sources of information about at-risk standard data elements is needed, including updating DICOM PS3.15 E.1 as new standard data elements are added and errors are found.
2. Encourage the development and sharing of de-identification software that can automatically ingest sources of information about at-risk standard data elements, including DICOM PS3.15 E.1, such that the most recent information is always used.
3. Continued maintenance of sources and use of information about known safe private data elements is needed, including updating DICOM PS3.15 E.3.10, as well as creation and maintenance of reliable sources that can be ingested automatically by de-identification software. This approach should be extended to include documentation and classification of proprietary fields used in proprietary WSI formats.
4. Further research is needed into improving the reliability of automated means of analyzing unstructured text (including text found burned-in to pixel data) to detect direct and indirect identifiers, and sensitive attributes, such that they may be automatically redacted without human intervention or review, such that only minimal QC is required.
5. Further research is needed into means of quantifying the reliability of the de-identification process, whether manual or automated, such that what is intended to be removed or replaced is actually removed or replaced, and how to express this in a meaningful and understandable manner, such as by one or more "scores". This is relevant both for the consumer selecting a process, as well as comparison of different processes, such as in a competition or challenge.

6. There is a need to create and promulgate additional test, training, and evaluation data sets for the de-identification process, software and algorithm improvement, with known correct answers. These should include direct and indirect identifiers, and sensitive attributes that are as realistic as possible.
7. Further research is needed into the performance of de-identification processes, software and algorithms on actual patient images and accompanying data elements, such that empirical quantification of actual re-identification risk is possible, and indeed assessment of the actual potential for harm given the additional information disclosed by successful re-identification. It is expected that some form of distributed experiment across large academic medical centers with the appropriate ethical controls and approvals would allow this to be conducted without the need for exposure of actual identifying information beyond each participating sites' control.
8. Further research (including thought experiments, modeling and simulations, and empirical experiments) should be performed into quantifying the actual incremental re-identification risk of potentially reconstructable facial information in head and neck cross-sectional images, to realistically assess the need for restricted access instead of public sharing, so as to balance that risk against the diminished utility of limiting access to, or de-facing such images, especially for head and neck cancer.

## 1.8 Background

The widespread routine public sharing for research purposes of medical images of patients obtained during clinical practice or clinical trials is recognized as a high priority. The ability to replicate and reproduce results is a core element of modern scientific practice [O'Connor 2017] [Obuchowski 2015]. The principles of Findability, Accessibility, Interoperability, and Reusability [Wilkinson 2016] are now *de rigueur*. An updated NIH policy emphasizes sharing [NIH 2020]. The commitment to publicly share federally funded scientific data in the US has recently been reemphasized [White House 2022]. Secondary reuse of image data has long been practiced in the cancer community [Clark 2013], and especially encouraged for research into computer aided detection (CAD) [Clarke 2001] and radiomics [Aerts 2014].

Demand for imaging data has exploded with the recent popularity of research into artificial intelligence applied to medical images [Kohli 2017] [Willemink 2020] [Kanakaraj 2022]. Existing archives of high quality curated images can be reused and extended for machine learning applications [Prior 2020]. A high volume of data from multiple different sites and systems is required for training and testing, and is essential for external validation to achieve generalizability [Park 2018]. Public or open-access sharing has been recognized as an opportunity to promote novel research [Bertagnolli 2017]. Accessible images have been described as a public good and have demonstrated their utility in this form [Batlle 2021b]. The popularity of challenges as a means of crowd-sourcing knowledge has increased the prominence of public sharing [Saez-Rodriguez 2016]. Barriers remain to widespread acceptance of image sharing for re-use [Bosserdt 2019] [Prevedello 2019].

It is essential to find a balance between protecting the rights and well-being of the individuals whose images are contributed, with the retention of scientific utility for research for the public good. The sharing of images publicly presents a particularly high bar for the prevention of re-identification, since once data are released into the wild, retraction is difficult, and any harm done may be difficult to quantify or recompense.

Alternative methods (than data sharing) of privacy-preserving research interactions have been proposed and are active areas of research. Federated learning can be applied to images [Sheller 2019]. Generative Adversarial Networks (GANs) can produce images that are helpful for training deep learning networks, and can generate large datasets that may more fully sample the distribution of the true population than real data (with or without conventional data augmentation) [Osuala 2022] [Osuala 2023]. These are possible ways to reduce (though perhaps not eliminate) the need for sharing de-identified images, but for the time being and the foreseeable future, conventional image sharing remains the mainstay of many research projects.

This report reviews the technical aspects of processing medical image data from clinical sources to prevent re-identification yet preserve scientific utility, in the context of the most challenging scenario, unrestricted public sharing. It will provide best practices as well as recommendations for future work in this area. The lessons learned are also applicable to less challenging scenarios that entail data use agreements, and registered, limited, or restricted access.

## 1.9 Goals

The Medical Image De-identification (MIDI) Task Group (TG) aims to:

- reach consensus on best practices for image de-identification for the purpose of public sharing of medical image data for secondary re-use,
- provide input to National Cancer Institute (NCI) Center for Biomedical Informatics and Information Technology (CBIIT) and other Institute or Center (IC) activities related to medical image de-identification,
- make recommendations on criteria and resources for performance evaluation of image de-identification tools and processes,
- make recommendations on appropriate quality control, assurance, and management efforts to be applied during de-identification,
- provide guidelines for image de-identification using manual and automated, local, centralized and cloud-based approaches, taking into account their portability and scalability

## 1.10 Scope

### 1.10.1 Modality

The primary interest of NCI in this context is cancer research involving images and image-related information. Human imaging is of great interest, including both in vivo (clinical) and ex vivo (biospecimen) imaging modalities.

For clinical imaging, traditional radiology modalities for cross-sectional structural (CT, MR) and functional or molecular (NM, PET) as well as cancer-specific projection X-Ray (chest and skeletal X-Rays, Mammography, Breast Tomosynthesis) are of primary interest. Other radiological modalities such as Ultrasound (US), particularly dynamic contrast enhanced US (DCE-US) may also be of concern. Photographic modalities (such as skin photography and dermoscopy, and still-frame or video gastrointestinal endoscopy) are of interest, as are specialized methods such as Optical Coherence Tomography (OCT).

Specimen imaging includes photography of gross tissue specimens, brightfield and fluorescence slide microscopy of frozen or fixed tissue section specimens stained with chemical dyes, antibodies, or molecular probes (including immunohistochemistry, immunofluorescence, in situ hybridization, and fluorescence in situ hybridization), electron microscopy of tissue section specimens and various modalities for ex vivo imaging of thick tissue specimens in three dimensions, including fluorescence confocal or two-photon excitation microscopy, micro-optical coherence tomography (µOCT), and micro-computer tomography (µCT).

Therapeutic modalities, especially external beam radiotherapy (RT), are planned and quality controlled using images. The entire gamut of image and image-related RT artifacts are also within scope (including RT plans, structure sets and dose volume histograms (DVH)).

Though animal imaging for pre-clinical research and veterinary applications is potentially within scope, these fields do not raise significant privacy concerns, though de-identification of researchers, handlers, owners and patient identifiers in xenograft tumor models described in the image metadata may be a concern.

Image-related information includes reports, documents, and annotations in various formats.

### 1.10.2 Purpose (Context)

This report will address only technical matters related to de-identification of images and associated data that are intended to be publicly shared without any restriction on their re-use and re-distribution. Though in practice such sharing may be accompanied by data use statements, licenses, agreements, or even criminal prohibition of re-identification attempts [Phillips 2017], it will be assumed that the actual ready availability of the data will preclude the effectiveness of such protections in the face of a nefarious adversary.

Similarly, matters of ownership, need for or revocation of consent, and legal, ethical, or moral issues related to the secondary use (including sale) of data are also out of scope.

## 1.11 Terminology

### 1.11.1 De-identification, Anonymization and Pseudonymization

Some confusion exists with respect to terms such as de-identification, anonymization and pseudonymization. Historical and geographical usage has diverged, and various standards and publications conflict with each other, as well as with applicable legislation in different jurisdictions.

For clarity and simplicity, this report will use the term "de-identification" exclusively and eschew "anonymization" or "pseudonymization".

Further, we will define "de-identification" to mean the removal of information that might allow re-identification. In practice we will use de-identification by reference to some objective standard, either in terms of a list of what information is to be removed, or some quantitative measure of permissible residual re-identification risk. Removal in this definition is intended in an inclusive sense and may include other forms of alteration.

This definition is not inconsistent with, and subsumes:

- the definition of de-identification in the ISO standard on pseudonymization [ISO TS 25237:2008], *"any process of removing the association between a set of identifying data and the data subject"* as well as the very similar definition therein of anonymization, being a *"process that removes the association between the identifying data set and the data subject"*;
- the use of the term in the HIPAA Privacy Rule [OCRa], which specifically defines a standard for de-identified protected health information to be (§164.514(a)) *"health information that does not identify an individual and with respect to which there is no reasonable basis to believe that the information can be used to identify an individual is not individually identifiable health information",*
- the use of the term in DICOM PS3.15 [NEMA PS3.15 E.1], in which *"profile and options support the de-identification of datasets to prevent leakage of individually identifiable information, for reasons of privacy",*
- the use of the term in the IHE ITI De-Identification Handbook [IHE 2014], which defines de-identification as *"any process that removes the association between a subject's identity and the subject's data elements",*
- the use of the term in various publications from US NIST to mean *"records that have had enough PII removed or obscured such that the remaining information does not identify an individual and there is no reasonable basis to believe that the information can be used to identify an individual"* [McCallister 2010], or *"removal of personal information from data that are collected, used, archived, and shared"* [Garfinkel 2015] [Garfinkel 2022].

Our use of the term "de-dentification" applies regardless of the method of de-identification and is not restricted to only rule-based approaches [Chevrier 2019].

It is understood that the term "anonymization" is sometimes used to imply an irreversible de-identification, and/or a complete de-identification. However:

- whether or not to retain a means of tracing the original subject, for whatever purpose, is somewhat independent of the means itself, whether it be through a map of pseudonyms, reversible encryption, or any other means, and whether or not keys or maps have been retained or discarded [McCallister 2010], and hence is separable from the de-identification process;
- a complete de-identification (in the alternative, a zero residual risk of re-identification) is impractical if not impossible [Finck 2019] [EDPS 2021] [Seastedt 2022].

The European GDPR [EU 2016] makes use of the terms anonymization and pseudonymization [Kogut-Czarkowska 2021], and not de-identification. It excludes anonymized records from its scope [Finck 2019] [Nasseh 2020]. GDPR considers pseudonymized records, those that require additional information to be re-identified, still to be a source of risk, and requiring of consent [Mostert 2016] [Peloquin 2020] [Vokinger 2020]. For the sake of argument, we shall consider 2016/679 (1)(26) *"personal data rendered anonymous in such a manner that the data subject is not or no longer identifiable"* to have been "de-identified" per our definition; the same recital addresses a standard for de-identification in the sense of requiring accounting for *"all the means reasonably likely to be used"* [Lewis 2019]. We recognize that some authors, particularly in a European or UK regulatory context, will continue to use alternative terminology [ICO 2012] [Sariyar 2016] [Elliot 2016a] [Batlle 2021a] [White 2022]. De-identification as we use the term corresponds to factual anonymity, as distinct from absolute or formal anonymity [Forschungsdatenzentrum Anonymity], or functional anonymity, which is data release environment dependent [Elliot 2018].

In this report, the term "anonymization" will not be used further.

This treatment of terminology is consistent with that of other publications on the subject [HEISC 2015] [Parker 2021a] [van Ooijen 2021].

"Pseudonymization" will only be used to mean the process of replacing direct identifiers with pseudonyms, in the sense of the plain and ordinary meaning of a pseudonym being *"a fictitious name used ... to conceal ... identity"* [Pseudonym].

This definition is not inconsistent with, and subsumes:

- the definition of pseudonymization in the ISO standard on pseudonymization [ISO TS 25237:2008], *"particular type of anonymization that both removes the association with a data subject and adds an association between a particular set of characteristics relating to the data subject and one or more pseudonyms"*
- the definition of pseudonymization in the European GDPR [EU 2016], *"the processing of personal data in such a manner that the personal data can no longer be attributed to a specific data subject without the use of additional information, provided that such additional information is kept separately and is subject to technical and organisational* *measures to ensure that the personal data are not attributed to an identified or identifiable natural person"*

In our discussion we will also distinguish reversible pseudonymization, in which *"the pseudonymized data can be linked with the data subject by applying procedures restricted to duly authorized users"* [IHE 2014]. The key feature being the secure maintenance of a relationship that allows identity to be reestablished [ENISA 2018]. We will not consider different degrees of pseudonymization [Hintze 2018].

A "collection" is defined to be a set of images and related information to be de-identified and released as a group, which typically consist of images belonging to a cohort of patients related by a common disease (e.g., lung cancer), image modality or type (MRI, CT, histopathology, etc.) or research focus (after TCIA). It will be used in preference to the more general "data set", which is encountered in the literature when referring to a set of records of individual subjects to be de-identified, and which represents a sample of a real-world population.

### 1.11.2 Identifiers and Attributes

In this report we will concern ourselves with the identifiability of any individual, whether it be a patient in the role of a research subject, or any other individual related to a collection being processed. This will encompass all "personal data" in GDPR *("related to an identified or identifiable natural person"*) [EU 2016], Personally Identifiable Information (PII) [GAO 2008] *"any information about an individual"* [McCallister 2010], and Protected Health Information (PHI) (45 CFR. § 160.103) *"individually identifiable health information"* [OCRb].

The term "identifier" will be used with its plain and ordinary meaning, *"a person or thing that establishes the identity of someone or something"* [Dictionary.com Identifier], and in this context will be limited to a variable that is an identifier of an individual natural person.

The term "direct identifier" will be used for variables that can be used to uniquely identify (be tied directly to) an individual, simplified from [El Emam 2015]. Examples would be national identity numbers (such as the US Social Security Number (SSN)), medical record numbers (MRNs), and driver's license numbers, which are nominally assigned to an individual. Other identifiers such as names, full addresses, and telephone numbers, which though not always unique to an individual, are also generally treated as direct identifiers, to the extent that they serve only for identification purposes, do not provide utility if preserved, and hence are routinely completely removed during de-identification. Date of birth, though also obviously not unique per se, is also often treated as a direct identifier and removed, in lieu of retaining age.

Whereas an "indirect identifier" is one that can be used to reidentify a subject in combination with background knowledge (something else an intruder might know) [El Emam 2015] [EDPS 2021]. Examples would be sex, age, or geographical location. Indirect identifiers are also sometimes referred to as quasi-identifiers [Dalenius 1986] or pseudo-identifiers.

Other than identifiers, there may be variables in the data whose disclosure may have undesirable consequences; these are "sensitive attributes". These may be the subject of "attribute disclosure" as distinct from "identity disclosure" [Sariyar 2016].

In this report, "metadata" will be used in a general sense to apply to any information that is not image pixel data but is in some way related to (about) the pixel data, including metadata encoded within the same file, embedded within compressed bitstreams that are encoded as pixel data, as well as accompanying metadata that may be in separate files but is intended for conjoint release with the images. This image metadata includes reports, documents, and annotations (about the images), which in another context might be considered to have data and metadata in their own right.

## 1.12 Substance

### 1.12.1 What is in an image?

To a lay person, a digital image is something encountered on the web or created by a camera, possibly within a mobile communication device (phone). As such the essential characteristic of an image is that it is a picture that contains visible information including people, their faces and other anatomic parts, and their environment. It is well understood that such content may be compromising [DMCA 2022]. It is becoming increasingly well understood, even by non-experts, that digital images may contain other information embedded within them, usually in the metadata but possibly also hidden within the pixels of the image. This information may be of utility to the user or used for other purposes, including tracking the creators and users of the images.

Like consumer images, medical images also consist of pixel data and associated metadata embedded in the file or conveyed in close association with the image and may contain direct or indirect identifying features or attributes as well as sensitive content [Garfinkel 2015].

### 1.12.2 Format

Consumer images, particularly those from phones and cameras, are generally encoded as JPEG [ISO10918-1] compressed bitstreams, usually with a JFIF header (marker segment) [ISO10918-5], though other formats are sometimes used [WebP]. Increasingly, more detailed metadata is encoded within JPEG images using the EXIF format [CIPA 2012], even by consumer phones and cameras. Scientific users may use other more specialized formats, such as TIFF [Aldus], which have evolved from computer imaging and desktop publishing applications. Specialist photographers may use so-called Camera RAW images [Wikipedia RAW], or DNG format images [Adobe DNG] to gain more control over the post-acquisition image processing. All these formats are important to consider for medical imaging de-identification because some medical images originate as such photographs, and dedicated medical formats may embed or encapsulate images encoded in such formats, especially JPEG. In addition, it may not be obvious that metadata with privacy implications may be embedded in the metadata of such formats, including the identity and personal information of the photographer, of their device, date and time and location of the acquisition, even including GPS-derived coordinates [Harvey].

That said, the dominant file format in clinical practice for radiology, cardiology and radiotherapy image acquisition and distribution is DICOM [NEMA DICOM]. In addition, there is a long history [Dayhoff 1992] of managing all forms of clinical medical imaging, including those from so-called "visible light" specialties, in DICOM format, since this is both technically feasible and advantageous, and allows for a single platform for enterprise-wide image management [Clunie 2016a].

De-identification of DICOM images and related information will be the primary focus of this report, and the primary emphasis will be on images of radiological origin, though the principles are generally applicable to all formats and applications.

It is recognized that individual researchers may elect to transform acquired radiology images from DICOM to some other format, to perform specific processing and analysis tasks. This may be done before or after de-identification. Since preserving much of the rich metadata included within DICOM files is important to maintain the utility of the images for as broad a range of unanticipated secondary re-use purposes as possible, it will be assumed that public sharing of de-identified DICOM data and related information is the primary goal. This goal is the most demanding from the perspective of preservation of subject's privacy. The general principles are equally applicable to pixel data and metadata shared in research formats.

The format for biospecimen image encoding is less well standardized. It has always been possible to encode single channel, multichannel, and color images of any type in DICOM Secondary Capture images [NEMA 1993] and even in the ACR-NEMA format that predated DICOM [NEMA 1985]. A DICOM standard for gross and slide microscopy has existed since 1999 [DSC Sup 15] and for WSI since 2010 [DSC Sup 145]. Given the very limited adoption of digital technology by clinical anatomical pathologists outside research applications, coupled with the relative paucity of regulatory approvals for slide scanning medical devices [Flotte 2018], manufacturers have not yet prioritized interoperability and standardization. Proprietary formats and de facto academic standards have proliferated in the research community, many based on semi-professional consumer formats like TIFF [Aldus]. Some, such as OME-TIFF [Goldberg 2005] [Linkert 2010], have been augmented with application-specific metadata encoding schemes. Enumerating the many WSI formats is beyond the scope of this paper. Within scope are general de-identification issues related to microscopy and slide overview and label pixel data, as well as specimen- and slide-related structured and unstructured metadata. They will be addressed to the extent that such issues are format-agnostic, specific to the biospecimen microscopy imaging process, and are similar to, or need to be distinguished from, DICOM radiology image de-identification.

### 1.12.3 What Needs to be De-Identified?

An informal taxonomy of the information of potential privacy concern includes:

- metadata
    - structured

- text
        - numeric
        - binary
    - unstructured
        - text
        - binary
- pixel data
    - content
    - features
        - faces
        - internal and external genitalia
        - distinguishing or unusual structural characteristics
            - direct (iris, retina, fingerprint)
            - indirect (shape of skull, pelvis, birth defects)
        - modifications (tattoos)
        - physical objects (implanted medical devices, dentistry, jewelry)
    - embedded metadata
        - burned in (or photographed) unstructured text
            - regular font (different size, emphasis, contrast, color, orientation)
            - cursive, including hand-written notes
            - logos (e.g., institutional)
        - hidden (watermark, steganographic)

These items include information of potential concern with respect to confidentiality (identity leakage) as well as sensitive attributes. E.g., the external genitalia are potentially an indirect identifier, i.e., a surrogate for the structured encoding of sex, as well as being potentially sensitive from a patient, staff, or public policy perspective, whereas the shape of the pelvis is possibly the former but less likely the latter.

A robust approach to de-identification for public release needs to account for all these items during risk assessment, though in practice many may not be considered as significant concerns, depending on the type and source of the images.

For example, it may be deemed extremely unlikely that steganographic information containing recoverable identifying information would be hidden within genuine clinical images submitted from reputable sites that are not known to be trying to deliberately undermine or discredit the process; hence detection of it would be excluded from further consideration. If watermarking with embedded identifying information was to become commonplace as a means of detecting tampering, de-identification would likely be rendered impractical unless the watermarking mechanism was fully reversible [Zhou 2021a].

Orthogonal to this taxonomy are other dimensions of classification of identifiers and attributes that may be of specific concern for removal or retention. Dates and times, for example, may be a concern for re-identification and preservation of utility, and will be discussed separately.

Note that all available information in the collection will be considered as candidates for de-identification. Matters with respect to what it is permitted to disclose in a particular setting (e.g., per the subject's consent, the IRB protocol, etc.) are out of scope.

## 1.13 Overview of De-Identification Approaches

### 1.13.1 How to Perform De-Identification

At a very high level, de-identification approaches are usually categorized as:

- rule based, or
- statistically based.

A simplistic illustration of this classification (after [El Emam 2013a] [Garfinkel 2015]) is the provision of two approaches in the US HIPAA Privacy Rule [OCRa] [OCRb]:

- one of which (45 CFR § 164.514(b)) lists a set of identifiers to be removed ("safe harbor" method, henceforth, the HIPAA Privacy Rule Safe Harbor (PRSH) method), and
- the other (45 CFR § 164.514(a)) that requires that a person with the appropriate expertise *"determines that the risk is very small that the information could be used ... to identify an individual"* ("expert determination" method).

In the imaging de-identification community, historically a simple rule-based approach has often been used, if for no other reason than the DICOM standard explicitly specifies an information model and corresponding attributes, encoded as structured data, which identify the patient and describes their characteristics in general and at the time of image acquisition. All of the Attributes of the DICOM Patient Module and Patient Study Module, and in particular commonly occurring data elements such as Patient's Name and Patient's ID, have always been obvious candidates for removal or replacement for de-identification [Kallepalli 2003] [Hackländer 2005]. It is logical to determine which DICOM data elements contain information that corresponds to HIPAA PRSH identifiers [Fetzer 2008].

By contrast, in the microdata community, a statistical approach is often used. "Microdata" is a term that was coined, initially in a census context, to describe *"records with information about individual data subjects",* even when individual direct identifiers are not present, for which there is a recognized non-zero risk of statistical disclosure [Alexander 1978] [Cox 1979] [Elliot 2018]s [EDPS 2021]. Microdata is distinct from aggregated information. Clearly, the metadata associated with (or derivable from) medical images constitutes microdata, when the images are attributed to individual human subjects, and when those subjects' characteristics are described. Though medical images do not generally contain a lot of information about the patient, as opposed to the image acquisition, in some cases it is important to retain "patient characteristics" (e.g., age, sex, weight and height) with the image to perform some types of image processing to generate physiologically meaningful quantitative results (e.g., FDG-PET

SUV). Accompanying relevant patient-specific information about medical history, therapy and outcome also needs to be considered.

Statistical disclosure control (SDC) is an approach to reducing the risk of re-identification of microdata. SDC can be considered as a problem in managing the risk of uniqueness in the population from which the microdata sample is drawn [Bethlehem 1990] [Greenberg 1992]. Depending on the size of the collection, the information retained, and if (or how) that information is fuzzed or binned or has noise added, the uniqueness of the included individuals within the population and the risk of re-identification, may need to be quantified. Considerations may be general, when a de-identification process is deployed, and collection-specific, when a well-established process is re-used for a new set of records, perhaps with different characteristics. The medical imaging de-identification community has been seemingly unaware of the statistical disclosure risk of patient descriptive metadata, beyond application of the HIPAA PRSF requirements. Only recently has attention been drawn in the literature to this aspect [Lien 2011] [Willemink 2020] [Parker 2021a].

In this report we will explore the premise that both rule-based and statistical approaches should be used in combination. The latter can only be elided when the rule-based approach removes all direct and indirect identifiers as well as sensitive attributes from the embedded image metadata, image pixel data, and all accompanying data. This may often be possible but may diminish the utility of the released data. Very commonly, some metadata containing indirect identifiers is released and the statistical risk ignored, which we suggest may sometimes be unsatisfactory.

We reiterate that both rule-based and statistical approaches need to be considered with respect to the nature of the information, irrespective of how it is encoded. Specifically, the same principles apply whether indirect identifiers are structured or buried within unstructured text, whether they are in header metadata elements or burned in to the pixel data, or whether they are derivable image features (such as sex from genitalia or other morphology).

Motivated intruder attacks are not considered an approach to de-identification per se, as distinct from evaluating the effectiveness of de-identification, so will be considered under the heading of quality control.

## 1.14 Rule-Based De-Identification

The simplest, most commonly used, and certainly easiest to implement means of applying a rule-based approach in an imaging context, is to remove or replace certain data elements in the DICOM image "header".

Like data encoded in other standards of the same generation (such as TIFF [Aldus]), ACR-NEMA and later DICOM datasets (whether transmitted on the network or stored in files) consist of a binary stream of data elements encoded as tag-value pairs, wherein the tag ostensibly defines the meaning of the data element [Bidgood 1992]. In ACR-NEMA, the encoding was a flat list of data elements sorted by ascending tag number. In DICOM, the possibility of nesting lists of data elements into sequences was added. This complicates de-identification, since recursive processing may be needed [Eichelberg 1995]. DICOM also added an object-oriented information model. This introduced the formality of documenting the data elements as being attributes of information entities (such as the Patient or Study), and grouped attributes into modules of those entities [NEMA PS3.3]. This organization is not reflected in the actual encoded data, however. There is still a need to address the individual data elements during de-identification, whether it be by enumerating them independently or reconstructing the information model. The numeric tags that represent each data element are defined by the standard in a data dictionary [NEMA PS3.6], which also assigns human-readable names (and keywords) to each tag.

Given such a structured set of data elements, there are two fundamentally different approaches possible, based on a list of elements or elements with certain characteristics:

- remove everything except that which needs to be retained (a permitted list, allowlist or whitelist)
- remove that which is known to be unsafe (a restricted list, denylist or blacklist [Etymology Blacklist])

Some combination of the two approaches may be applied for different categories of information.

With either approach, consideration needs to be given to elements whose status is unknown. This may include private data elements created by manufacturers, or recently added new standard data elements of which the software is unaware. On the one hand, unrecognized elements may be potentially unsafe; on the other, they may be critical to preservation of scientific utility.

Rules may be simple lists of data elements to remove, retain, or modify in some manner. Or they may be more complicated, and specify behavior based on more than one data element at the same time, or be conditional on some internally recognizable constraint (such as SOP Class or other indication of the kind of image), or externally supplied predicate.

An obvious approach to preparing a list of DICOM data elements that present a potential disclosure risk is to manually review the standard Data Dictionary or Modules for Data elements described as being:

- direct identifiers, for example, the Patient's Name, Patient's ID, or
- indirect identifiers such as Patient's Age or Patient's Sex.

In some cases, the selection process may be guided by what is routinely encountered in practice. Many DICOM data elements are unused by most implementations or are specific to circumstance. In our experience, this selection task is non-trivial and difficult to perform well. Indeed, it is hard to imagine that an individual researcher will have the time to perform the initial analysis, much less the time to routinely review the local practice when new elements are added to the DICOM Standard, existing imaging modalities receive software updates, or new imaging modalities are added to a local or remote site. Instead, we recommend using an existing list or rule set from an authoritative source.

### 1.14.1 Profiles for Rule-Based De-Identification

An early attempt to add some formality to a rule-based list of offending attributes approach was the addition in 2002 of a De-identification Profile to DICOM [DSC Sup 55]. This included a relatively short list of identifiers and descriptors of patients, other persons, devices, and institutions as well as unique identifiers (UIDs). Some publications explicitly referenced the supplement that added this profile when describing their de-identification process [Moreira 2012]. Experience within the pharmaceutical research industry [Clunie 2007], with teaching files [IHE TCE] and in the academic research community [Freymann 2012] drew attention to the inadequacy of the short list. A much more comprehensive list was added to DICOM [DSC Sup 142] and continues to be diligently maintained [NEMA PS3.15 E.1]. The scientific literature is now starting to refer to this as an appropriate basis for de-identification [Patel 2014] [Kundu 2020a] [Aiello 2021] [Yi 2021] [Lazic 2022] [Kondylakis 2023] [Kondylakis 2024], as are regionally-specific guides [MISA 2019].

Various authors have published their own lists of what to remove [Noumeir 2007] [Fetzer 2008] [Cosson 2012] [Aryanto 2012] [Newhauser 2014] [Aryanto 2015] or what to retain [Parker 2021b]. We are not convinced that any of these are sufficient or safe. Instead, we strongly recommend the use of the list specified in the DICOM standard itself [NEMA PS3.15 E.1]. We recognize that the DICOM-supplied list may not be perfect, and its comprehensiveness may be daunting, but we strongly disagree with those who contend that it is impractical, though sympathize with their concerns [Parker 2022]. It is the most complete list that we know of that specifies what is potentially unsafe without removing more than is necessary, is subject to review and revision on a regular basis, is maintained as new standard data elements are added for new applications, and can be implemented with automated tools. Automatic conversion of the DICOM standard list into machine-processable rules has also been described [Onken 2009].

The design of the DICOM de-identification profile is based on the premise that, with few exceptions, the data elements contain what they are supposed to contain. This has been found to be a reliable assumption, except with respect to plain text strings such as descriptions, to which human operators have edit access. Accordingly, the default behavior of the profile is to require removal or replacement of all such textual descriptive attributes. The default also specifies removal or replacement of all the data elements that are known to be direct or indirect identifiers. This includes indirect identifiers of not only the patient, but also of any human involved in the imaging process, as well as identifiers and characteristics of the institution and the imaging device. These could all be construed as potential indirect identifiers. The default also includes removal or replacement of all dates and times. All private data elements are also required to be removed when applying the default profile. The default is intended to be readily applied by simple software using straightforward mechanical means without significant analysis.

From the already conservative basis that is the default profile, additional protection can be gained if additional named options to remove more information are implemented. These options relate to removal of burned in identification within pixel data or bitmap or vector graphic overlays, as well as identifiable features in images such as facial information; both will be discussed in detail separately. It is understood that such image processing may be hard, and sometimes impractical, unnecessary or may compromise utility, hence the additional protection is specified as separate options.

Other named options to the DICOM de-identification profile address the matter of selectively retaining more information, rather than removing it. These options include retention of patient characteristics, device or institution identification and characteristics, as well as so-called "known safe" private data elements. Additional options address the treatment of unique identifier (UID) data elements, which have a structural significance in preserving referential integrity, and dates and times, and the retention of textual descriptions that have been "cleaned" (analyzed and identifying information selectively removed).

Each of the named options for retaining information addresses a specific class of use cases. For example, retention of patient characteristics such as age, sex, and weight of the patient may be essential for quantification of the images when the metric involves the volume of distribution of an agent, such as a PET radiopharmaceutical. Device or institution identity or characteristics may be necessary for stratification of results to consider the quality control of images or generalizability of results. For simplicity of implementation, related data elements are grouped together given that a use case may require such sets. We recognize that this grouped approach may not suffice when statistical disclosure control provisions come into play, especially with respect to the patient characteristics being potential indirect identifiers. It reflects the immaturity of the application of SDC techniques to DICOM images that there is as yet no coded and structured means of describing what changes have been made on a statistical basis, or to indicate that selected data elements have been subject to perturbation rather than removal or replacement individually or as a group. The standard may need to be extended to better describe SDC approaches and document the techniques and parameters used.

The institution information, particularly its address, may be of concern to the extent that it provides geographic location information about the patient, albeit relatively crude, which is a known source of increased risk [Sweeney 1997]. Depending on the type of facility though, institution identity may not be a good proxy for location, since referrals may be from a distant location. Leakage of institution identity may occur through metadata about the entire collection (e.g., a set of images from a single facility).

Device characterization, if not identification, may be important for image analysis and quality control. Knowledge of the device used may be important from a bias perspective (e.g., patient weight might be correlated with model; older models may be in poorer neighborhoods). There is a theoretical possibility that the identity of the institution could be inferred from exotic acquisition characteristics, such as extremely high MR field strength or use of novel technology.

The public availability of device databases to match serial number or model or manufacturer to institutions is only a theoretical concern at present. Even if specific device characteristics are removed, they might be derived from the characteristics of the pixel data. It is common practice to retain manufacturer and model data elements but not serial number, if only to allow for the demonstration of generalizability of AI models. The manufacturer and model may also be implicit from the pattern of standard data elements present, as well as specific values of protocol names, study and series descriptions, not to mention any private data elements (which are otherwise nominally safe).

Some patient characteristics such as age may require special treatment based on jurisdictional rules that have a statistical basis, even in the absence of a formal statistical methodology being used. E.g., the HIPAA PRSH method requires ages over 89 years to be aggregated [OCRb]. See the later discussion on the topic of age.

The DICOM profile does not explicitly categorize data elements as being direct or indirect identifiers or sensitive attributes. All direct identifiers of the patient and other human persons are always removed. Some of the named options allow for retention of groups of potential indirect identifiers. Sensitive attributes (to the extent that there are any defined in DICOM in a structured manner and recognizable as such) are flagged to be removed or cleaned, for example diagnosis.

The DICOM profile also addresses the encoding of structured information that is contained within reports, documents, and annotations. The option to clean structured content specifies a similar rule-based approach to the baseline profile, but by addressing coded concepts rather than traditional DICOM data elements. The coded names of the name-value pairs that constitute the content items of DICOM Structured Reports (SRs) and that pose a potential risk for identity leakage are enumerated. This list only addresses coded concepts found in standard DICOM SR templates, however, and should not be considered exhaustive. If no effort is made to clean structured content, it needs to be completely removed or replaced with dummy content [Noumeir 2007] [NEMA PS3.15 E.1].

### 1.14.2 Configurability of Rule-Based Implementations

Some tools that perform DICOM de-identification are configurable, either with a script or configuration file [RSNA MIRC CTP] [Google De-id] [Veldhuis] [Suzuki 2007] [Vcelak 2011] [AnonMed] [Microsoft DICOM], or through a user interface [González 2010] [Roddie 2016]. Such configuration mechanisms are powerful, but dangerous in the hands of those not intimately familiar with the problem. We strongly suggest that if configuration is to be undertaken, it should begin from a strong base, such as a full implementation of the most recent DICOM specification for de-identification.

Ideally, configuration should be limited to selecting one or more of the named options to retain or remove particular classes of information. Adding more data elements to the list to remove or replace is not unsafe but may compromise utility and compliance. Adding more data elements to retain without replacement may be unsafe and should only be undertaken after a careful risk assessment. Such an assessment should be thoroughly documented, subject to review by appropriate experts, and regularly reviewed. As a case in point, the CTP tool referred to earlier [RSNA MIRC CTP] is supplied with a script that implements the DICOM standard list [NEMA PS3.15 E.1]. Indeed, the DICOM standard list was developed in close consultation with the authors and users of CTP [Kirby 2011] [Freymann 2012]. It is important that the most recent CTP script reflecting the current standard be used, noting that the standard itself is updated five times per year.

### 1.14.3 Documentation of Rule-based De-identification

Regardless of how the de-identification tool is implemented and configured for a particular deployment, its characteristics and behavior should be thoroughly documented, either in terms of which profiles and options are used in the case of DICOM profile implementations, or down to the level of individual actions on individual data elements. DICOM defines so-called "Conformance Statements" as a means of describing compliance with the standard [NEMA PS3.2]. Recently the template for such statements has been updated and now includes components related to de-identification [Sup 209]. Prior implementations have been documented informally but with an appropriate level of detail [Hoffman 2019].

The DICOM standard contains specific elements in which the fact that de-identification has taken place (Patient Identity Removed), can be documented, and the de-identification process described (such as within De-identification Method Code Sequence), as well as specific standard coded concepts to be used for standard profiles and options. In addition, the de-identifying tool's characteristics may be recorded within the de-identified objects, such as by using the Contributing Equipment Sequence.

## 1.15 Statistical Disclosure Control

As described earlier, SDC is an approach to reducing the risk of re-identification of records with information about individual data subjects, which for our purposes includes images and accompanying metadata. Only a basic overview will be provided here, due to the specialized nature of the subject. The reader is referred to authoritative works on the subject for further information [El Emam 2013a] [El Emam 2013b] [El Emam 2015] [Arbuckle 2020].

It is important to maintain a realistic perspective when considering statistical techniques, since there is a tendency towards excessively academic or conservative techniques that render the data useless: "*While formal privacy models such as k-anonymity provide guarantees of protection, they can, at times, be too rigorous, leading to unacceptable levels of data utility in comparison to a rule-based policy*" [Xia 2015]. Mathematical models should be considered with respect to actual data available to an adversity: *"reidentification risk is situationally dependent and ... appropriate adversarial modeling may permit biomedical data sharing on a wider scale than is currently the case"* [Xia 2021].

A recent practical example of the application of formal risk assessment methodology to sharing EHR data (albeit not involving images), considering a variety of threats including information obtained from potential breaches, required a very modest suppression of individual records (<2%), subjects (<0.5%) and values of indirect identifiers (<2%), and an improvement over the HIPAA PRSH method [Thoral 2021].

The availability of statistical methods does not invalidate the use of the HIPAA PRSH method. Though concerns have been raised about its effectiveness [McGraw 2013], it has withstood re-evaluation [Benitez 2010] [Kwok 2011].

SDC techniques apply only to structured data. If data is not structured, whether it occurs within text or in text burned-in to images, it is first necessary to apply information extraction techniques to identify and separate structured data.

Note that all the structured data that is part of or linked to the data in the collection being released needs to be considered. This includes the DICOM header metadata, the information in accompanying spreadsheets or databases [Kundu 2021] [Kundu 2022] as well as the data in other collections that are directly linked (e.g., through common pseudonymous identifiers in different collections in different repositories).

Identity disclosure occurs when a person's identity is matched to an individual record by using values of the indirect identifiers, which have previously been defined, given that the direct identifiers have already been removed or replaced. Matching is also referred to as "record linkage" [Fellegi 1969]. Not only must an intruder establish a unique match on a combination of indirect identifiers in sample and population, but to be successful they must also establish that it a correct match [Elliot 1999].

There are other types of disclosure that are not re-identification, e.g., attribute disclosure, which we will not elaborate on further.

In the general case, it may be useful to know what knowledge will be gained by an intruder from any re-identification. Typically, this is some piece of information in the collection about a real individual, which is exposed by matching the record in the collection with a real individual.

*"Disclosure is a difficult topic. Even the definition of disclosure depends on the context. Sometimes it is enough to violate anonymity. Sometimes sensitive information has to be revealed. Sometimes a disclosure is said to occur even though the information revealed is incorrect"* [Lambert 1993].

In the case of collections for public release without additional controls, re-identification may be deemed to be problematic *prima facie* (regardless of actual harm).

SDC approaches are based on two principles, that:

- the risk (probability) of re-identification can be estimated, and

- the data can be modified to reduce that risk to an acceptable level and still retain some utility.

### 1.15.1 Re-Identification Risk

Statistical approaches to estimating and mitigating the re-identification risk can be further classified into two general approaches [Elliot 2000]:

- establishing some measure of uniqueness and its impact, or
- attempting to match individual records in a sample to a file representative of the population, i.e., mimicking an intrusion.

Note that "risk" in this context is the likelihood that a record in the released collection will be identified. This is distinct from other uses of the term "risk" in the medical device [Mahler 2022] and cybersecurity community [NIST 2012], or indeed more broadly [Crouch 1982]. It considers only the probability of the event and not the harm arising from its occurrence, i.e., not the severity of the hazard.

#### *1.15.1.1 Uniqueness*

Uniqueness is a core concept for modeling re-identification risk:

*"If a data release displays the information for an individual unique in the population, then an intruder will know that such an individual was included in the data base. An intruder who possesses matching data about a population unique has the potential to match his or her records against those in the data"* [Fienberg 1998].

If an individual is unique within a specified set, they may certainly be re-identified, i.e., the probability of re-identifying that individual is one (100%). If they are instead not unique, but a member of a group (equivalence class [Greenberg 1992]) that shares the same values for a set of indirect identifiers, then for a particular group size, the risk is the inverse of the group size. I.e., 1/2 for a group size of 2, 1/3 for a group size of 3, etc.

There will be variation in group size for different individuals in the collection. Some will be members of small groups, if not unique. Others will be members of large groups. How membership in groups differs contributes to the calculation of re-identification risk for the entire collection, e.g., maximum, or average, which depends on the choice of context (threat model) to consider.

##### 1.15.1.1.1 k-anonymity

Assigning the maximum risk of any group in the collection as the risk for the entire collection, for a group size of *k*, is referred to as *k*-anonymity, i.e., *"when the ... data do not allow the recipient to associate the released information to a set of individuals smaller than k"* [Samarati 1998]. This has also been described as the unique tuple of indirect identifier values as being *"hidden in a crowd of size k-1 other people"* [Aggarwal 2005b].

##### 1.15.1.1.2 Population Uniqueness

It has been implied that the group size is determined for the collection that is to be released. However, it is actually the risk in the population not the collection that matters, the collection being only a (typically very small) sample, and hence one needs to estimate the group size in the population. Estimating from the collection alone (such as by using *k*-anonymity) is extremely conservative and may result in the loss of a lot of utility if generalization is performed without considering uniqueness in the population rather than the sample. In other words, uniqueness in a sample is not synonymous with re-identifiability [Yakowitz 2011] [Yakowitz Bambauer 2015] [Sánchez 2016].

There are two approaches to determining uniqueness in the population:

- matching to a registry of the population, or
- estimating the characteristics of the population from the collection [Dankar 2012]

The difficulty of creating a perfect population registry, without which absolute certainty of a unique match cannot be guaranteed, has been observed [Barth-Jones 2017].

Despite the common perception that social media information represents a risk, practical efforts to attempt re-identification against social media data sources have yielded few matches with low confidence [Branson 2020].

Another factor not generally considered is the impact of healthcare enterprise security breaches that lead to data release; should all such records and the detailed information therein then be assumed to be generally available to any potential intruder?

#### *1.15.1.2 Threat Model*

There are two potential directions of attack:

- population to sample - someone who knows an individual, and knows they are in the collection, and may know lots of information about the individual, tries to match that information with indirect identifier values in the collection; also referred to as a prosecutor attack
- sample to population - someone selects a record in the collection, and attempts to match it to a real person; also referred to as a journalist attack; this requires some kind of register of everyone in population (e.g., voter registration database, or incomplete database such as might be constructed from social media); typically, the information available is not as detailed, and may be restricted to basic demographics and location

For risk analysis, the direction of attack matters. It affects both the probability, and which indirect identifiers to consider.

For publicly shared collections, the likelihood of attack is not considered, and is assumed to be a certainty, i.e., to have a probability of one [Elliot 1999] [El Emam 2013a]. Also, for publicly released data, the more conservative maximum rather than average risk is generally considered [El Emam 2013a].

#### *1.15.1.3 Selection of Indirect Identifiers*

Not every metadata element is an indirect identifier, and not every indirect identifier needs to be included in the risk analysis. Indeed, *"when the data contains a large number of attributes which may be considered [indirect] identifiers, it becomes difficult to anonymize the data without an unacceptably high amount of information loss"* [Aggarwal 2005a], yet retention of a large number of demographic attributes poses an unacceptably high risk [Rocher 2019]. Choice of which indirect identifiers to model, and which to ignore, depends in part on an informal estimation of the intruder's likely access to the information to perform record linkage (intruder power) [Xu 2008] [El Emam 2013b], and this is recognized as a potential weakness of the approach [Garfinkel 2015]. Indirect identifiers may also be highly correlated. Best practices from microdata community should be followed. There have been attempts to formalize the selection process [Elliot 2011].

Others have questioned this approach: "*Any separation between QIDs and SAs is essentially making assumptions about the adversary's background knowledge that can be easily violated, rendering any privacy protection invalid"* [Li 2012]; but their results may be too conservative to be of practical use.

Much of the re-identification and statistical disclosure control literature emphasizes precision of geographic locality [Elliot 1998] [Barth-Jones 2012] [El Emam 2011] as an important factor in increasing risk. To the extent that such information is often irrelevant for imaging use-cases, it can be omitted from inclusion in the public collection, thus may significantly reduce risk. Conversely, geographic location may be relevant from the perspective of bias and may be imputed with coarse precision from the identity of the institution [Kaushal 2020].

Patient descriptive characteristics such as age, sex, and race, as well as physical characteristics, may seem the most obvious potential indirect identifiers, each of which can be evaluated in terms of the likelihood of an intruder's access. Other less obvious attributes also need to be considered, such as diagnosis. The diagnosis may be explicitly present in the collection, either in DICOM data elements such as Admitting Diagnoses Description or Code Sequence, or in accompanying tabular form. Or it may be implicit from the very nature of the collection (e.g., a collection of brain tumor images). There has been research into the impact of diagnosis information on de-identification [Loukides 2010] [Loukides 2013]. The importance of preserving such characteristics has recently increased as their impact on evaluating bias and generalization in machine learning has become apparent [Schmitt 2021] [Driessen 2023].

Some have suggested that if only a small number of indirect identifiers are present in the collection (two or three), then no statistical risk assessment is required [Hrynaszkiewicz 2010]. Experts have concluded that there is no basis for such an assertion and that a proper analysis should always be performed [El Emam 2015].

### 1.15.2 Data Modification

To reduce the risk, the values of indirect identifiers for selected records can be generalized to produce more records that share the same values, and hence increase the group size [Domingo-Ferrer 2005]. Values (whether continuous, ordinal, or categorical) can be grouped together to produce more coarse categories, or values can be suppressed entirely [Kohlmayer 2015]. With global recoding, all attribute values are generalized to the same level of granularity, whereas local recoding allows for different levels of granularity for different records [Wong 2010]. It is not clear what to do with longitudinal data of the same subject, e.g., how to modify temporally affected indirect identifiers such as age when there is more than one record.

No changes to indirect identifier values are needed for those records whose group size is already sufficient.

#### *1.15.2.1 Utility Preservation*

As has been described, the goal is to balance reduction of re-identification risk with the preservation of utility [Kohlmayer 2015]. To optimize this tradeoff using a modeling approach requires that there be a suitable metric of utility preservation. These may be based on various generic measures of the data distortion of the original values, as well as various methods designed specifically for the purpose [Goldberger 2009].

An additional factor to consider is the need to choose perturbation methods that do not contradict features apparent from the actual content of the image pixel data. E.g., for images that make it obvious what the patient's sex is, one can remove but not change the sex in the metadata. Further, methods that account for the effect on image quantitation should be considered. E.g., generalization of height or weight values might affect computation of PET SUV.

More generally, one should consider that the metadata is present to assist with interpretation of subject-specific images, not for general analysis in its own right. Hence, some perturbation techniques that preserve the statistical characteristics of the set of records, but not the integrity of the individual record, may not be suitable. For that reason, generalization and suppression may be appropriate, but data swapping and the addition of noise may not.

For utility preservation in traditional (non-image) microdata record sets, statistical calculations performed on the data before and after de-identification can be compared. Similarly, image processing operations with objective (quantitative or categorical) outcomes, preferably incorporating metadata affected by de-identification, could be compared before and after. Such metrics could also be useful when scoring the effectiveness of de-identification methods, as discussed elsewhere.

#### *1.15.2.2 Threshold Choice*

There is always some risk of re-identification; it is never zero. So there is a need to define an acceptable threshold. There are precedents that have been used in practice and work well and have been confirmed by the failure of commissioned motivated intruder attacks. E.g., 0.09 is common choice. For more sensitive collections, a threshold of 0.05 may be used. The data quality difference between 0.09 and 0.05 is quite large. One approach to selecting a threshold is to make a comparable choice to the level of protection provided by the HIPAA PRSF [El Emam 2013a].

#### *1.15.2.3 Data Incompleteness and Divergence*

Data is usually not perfect, and this has an impact on re-identification attempts: *"differences between data sets ... and differences between data sets and the world ... reduce the likely success rate of matching attempts*" [Elliot 1999]. However, *"there is no meaningful way to adjust the uniqueness statistic to incorporate their effect"* [Elliot 2000].

How realistic are statistical methods, particularly when indirect identifier information is only partially complete, and/or indirect identifiers are multitudinous, which tends to lead to overly conservative estimates? There have been attempts to address these sorts of concerns [Xia 2021].

#### *1.15.2.4 Tools*

Several well-known freely available open-source tools are available that may be considered for medical image metadata re-identification risk analysis and mitigation. Commercial tools are also available. None are specific to medical imaging; their use would require:

- extraction of the image metadata into a tabular form
- analysis of the risks
- modification of the meta data until the risk was acceptable
- export of the modified metadata into tabular form
- replacement of the metadata in the images with the new values

There have been inventories of tools made in the past; they may not include the latest tools but still serve as a useful description of the features required [Fraser 2009].

##### 1.15.2.4.1 ARX

ARX is a relatively complete open source [ARX source] tool with a graphical user interface as well as an API. It allows for data ingestion, risk modelling using a large variety of parameterizable models, risk estimation, specification of a range of different generalizationh mechanisms, optimization of the application of those generalizations until an acceptable level of risk has been achieved with preservation of sufficient utility [Kohlmayer 2015], visualization of the modified data in comparison with original values, and export of the modified data. The tool is well documented [ARX] and described thoroughly [Prasser 2015] [Prasser2016a]. Practical examples of its use may also be found [Ratra 2022]. A video tutorial is also available [ARX 2015], though some of the user interface features are out of date. The approach to optimization used by the tool is separately described [Prasser 2016b].

#### 1.15.2.4.2 sdcMicro

The sdcMicro R package [sdcMicro] provides implementations of SDC methods to evaluate and anonymize confidential micro-data sets, and includes all popular disclosure risk and perturbation methods, is highly optimized to be able to work with large data sets, and contains reporting facilities to summarize the anonymization process [Templ 2015].

## 1.16 Operational and Deployment Considerations

The discussion to this point has been relatively abstract and has not considered how de-identification tools might be operationally deployed. The need for de-identification arises in many contexts ranging from an individual user's need to modify a small number of files for sharing, teaching or publication, a clinical trial site coordinator manually de-identifying individual patients' data before submission to a coordinating center, through very large-scale production deployments providing de-identified access to live clinical imaging data from multiple sites.

Since our focus is on the technical requirements of public data sharing, we are largely focused here on the end product, the de-identified collection, and the common features of what needs to be removed or replaced, as opposed to how to actually do it using a particular architecture. We recognize that such things as complying with regulatory and contractual requirements and scalability are important factors, however.

For example, on-site execution (in which identifiable data never leaves the site) may be necessary, as opposed to using an off-site (e.g., cloud) provided service [Deshpande 2014]. This may be mitigated by an appropriate contract or business associate agreement, coupled with whatever auditing and technical measures are necessary to trust a service provider's security, encryption-in-transit, and encryption-at-rest in addition to general security and access control mechanisms.

With the rapidly growing demand for large amounts of data, fully automated pipelined execution, including integration with real time retrieval and processing of production clinical images with de-identification, has been explored [Suzuki 2007] [Bland 2007] [Aryanto 2013] [Yi 2021] [Kathiravelu 2021] [Kanakaraj 2022]. One commercial vendor has referred to such a pipeline as a "data supply chain" [Dicom Systems 2017].

It is not uncommon in large-scale deployments to use an ensemble of tools in series. For example, a submitting site may perform its own de-identification (e.g., during extraction from the commercial PACS), the tool used for submission may perform a scripted de-identification on-site (e.g., using CTP [RSNA MIRC CTP]), then a dedicated team may use a tool like POSDA [Bennett 2018a]. The extent to which later stages catch more information of concern has not been documented to our knowledge. At a more granular level, individual tools may be assembled from multiple components that are repurposed, developed, maintained or distributed separately (e.g., pre-extraction of metadata, classification of unstructured text, OCR, remediation of headers, remediation of pixel data). Multiple such micro-services may be integrated to produce a cloud-based solution. In any case, any de-identification process that involves multiple steps or components presents a deployment challenge with respect to version control, since a change in any single component may compromise performance and previous validation testing.

## 1.17 Other Considerations

### 1.17.1 Structured Data Elements

A reliable understanding of what a data element means is a prerequisite for applying a rule-based approach or for analyzing the risk of indirect identifiers using statistical approaches. One needs to be sure that structured data contain only what they are supposed to contain, if one is to depend on the purported nature of the data element for de-identification decision-making. Except for text in descriptive structured data elements, this has generally proven to be the case. The strong typing of the Value Representation of DICOM data elements provides an additional level of reliability, as long as a check is performed to ensure that values encoded within data elements indeed conform to their requirements, especially for string rather than binary encoded values. Binary values that are numeric (whether integer or floating point) are self-constraining in this respect; though a numeric identifier could be encountered in such values unexpectedly, this is not generally considered likely. Binary large objects (BLOBs), on the other hand, may well contain undesirable information, and should not be considered "structured" in the absence of some definition of the format of the information therein. This is true both for standard data elements, such as those that contain encapsulated documents, as well as private data elements. Given the strong pressure on DICOM implementers to conform to the standard, this can generally be assumed to be the case, but nevertheless can and should be checked.

A similar approach can be applied to out-of-band metadata that is nominally structured, for example, as supplied in spreadsheets. Though CSV files do not generally contain type information, the proprietary formats or databases from which they are often derived often do, and there may be data dictionaries available in textual or formal syntax that can be used to determine what type checks to apply. For structured metadata acquired during the course of clinical trials, especially in therapeutic pharmaceutical trials, the PhUSE De-Identification Standard [PHUSE 2015] [Ferran 2015] for use with the Clinical Data Interchange Standards Consortium (CDISC) Study Data Tabulation Model (SDTM) may be consulted. It provides similar rule-based and general guidance for data elements in clinical data as DICOM does and provides alternative rules for different circumstances. Unlike DICOM, the PhUSE standard explicitly categorizes some data elements as direct or indirect identifiers.

### 1.17.2 Text in Structured Data Elements

Descriptive data elements that contain plain text, especially those that may be subject to operator modification, are challenging. This is more so when they often contain extremely useful content. Study Description and Series Description are particularly important in this respect. The values in these two data elements are often the only succinct explanation of the purpose and content of the images. They frequently contain information that is not easily derived from other structured data elements. It is no accident that Series Description remains to this day the primary data element used to drive hanging protocols, which direct how images are laid out on the screen in a PACS viewer [Morioka 2001]. They are also used for processing selection [Fadida-Specktor 2018]. Study Description is still used for prior relevant study matching and retrieval [Mabotuwana 2013].

Yet it is undeniable that operators sometimes insert highly inappropriate content into these fields. This sometimes includes patient and staff identifiers, descriptors, and contact information. Accordingly, a robust de-identifier that wants to retain the important content has no choice but to attempt to clean these values, removing inappropriate content and retaining what is useful. The rules-based approach defined in DICOM specifies that all structured descriptive data elements be removed or replaced unless a specific option to clean them is implemented.

Cleaning text in structured data may be implemented with data element specific techniques or treated as a special case of the general approach to cleaning unstructured text data. It may be addressed by automated heuristics, or other approaches to natural language processing, or if justified, by human review and editing. The two data elements (Study Description and Series Description) alone may justify the investment of significant human effort, even if all other descriptors are simply completely removed or replaced.

We note that the same text values may frequently recur within a series or a study, for an entire institution's set of data or even an entire collection or repository, particularly when they are populated automatically by the scanner from pre-defined protocols. Accordingly, it may be possible to configure (or learn) a set of acceptable strings or regular expressions that it is permissible to retain, while rejecting or sequestering all others. Or to put it another way, it is not usually necessary to make this decision individually for every word of each data element of every image being de-identified; historical experience can be re-used. For collections from multi-national sources, the descriptions may well be in the local language. This means that support of internationalized character sets, and thesauruses mapping common terms (such as those used for body parts) may be helpful. Just as individual data elements may be blacklisted or whitelisted, so too words or phrases or patterns occurring within text strings may be similarly enumerated.

The body part (anatomical region) is often described textually rather than using a formal code. The DICOM data element Body Part Examined may be of concern, since it is both extremely useful if present, yet has been observed to be subject to significant abuse by manufacturers, sites, and operators, though not necessarily posing a risk of identity leakage. Because it is a DICOM Code String (CS) value representation and not a plain text description, it is not included in the standard DICOM de-identification profile as requiring removal or cleaning. However, it may be a data element that is worth closer scrutiny, and it may also be worth considering excluding or remapping any non-standard values (e.g., those not included in PS3.16 Annex L [NEMA PS3.16 Annex L]), or even replacing the entire data element with the coded Anatomic Region Sequence data element [Towbin 2021]. Though this level of curation is beyond the scope of this de-identification report, it is worth mentioning, since diligent curation of the content of text fields and conversion to codes may preempt some de-identification issues.

Another peculiarity related to body parts is that they sometimes overlap with people's names. For example, Dr. Hand or Mrs. Leg. This is a consideration when de-identifying text strings such as Study Description by making use of a dictionary of acceptable or desirable words to retain.

On the subject of Code String (CS) data elements, as mentioned, these are generally considered safe (e.g., within the DICOM baseline profile), since they are usually machine-populated and not subject to human editing. It may still be worthwhile to monitor the values of such data elements as they are encountered to make sure they do not become a concern. Retention of their standard or manufacturer-provided values is generally important for utility preservation, since they may play an important structural role in defining the information object or convey important information about the acquisition technique or manner in which the data is encoded (e.g., Image Type). As for Body Part Examined, it may be worthwhile to check for values in CS data elements that are non-standard or not known to be safe proprietary values, and flag them for further attention.

Another consideration related to unstructured text string cleaning involves the use of known structured values as strings to match. For example, if the Patient's Name data element has been extracted and removed, its value can be searched for in unstructured descriptions, including those recognized by OCR of burned-in text [Vcelak 2019] [Pure Image], as well as that present in private data elements. However, in practice the benefit of this approach may be modest, since the supplied image data may already have undergone some limited de-identification before export from the original site, and obvious data elements like Patient's Name already removed or replaced with a pseudonym. At the original site though, such checks may be useful. Furthermore, the formal encoding of the Patient's Name data element as a Person Name (PN), with its embedded caret delimiters (e.g., "Smith^Jane") may be different from its use within other plain text strings (e.g., "Jane Smith"). These permutations, as well as phonetic variations and alternative spellings, can be accounted for and sought out. This also applies to other direct identifiers such as the medical record number encoded in Patient's ID, the Accession Number, and so on. Variations in formatting of date and time should also be taken into account, considering different delimiters, order of components, numeric versus string forms as well as less common nationally-specific forms.

Completely general approaches to de-identifying unstructured text (such as within clinical radiology reports) will be discussed later. In principle, some of these general methods can be reused for de-identification of unstructured text extracted from medical image headers, or from the pixel data by OCR. Examples of such tools are now available [Google De-id] [Vcelak 2019] [Pure Image] and being tested [Kopchick 2022]. It may be appropriate to provide special handling for some data elements of special significance (like Study Description, Series Description and Body Part Examined) as well as apply completely generally techniques that are sufficiently trusted to clean other textual descriptive data element values.

Heuristic techniques and regular expression driven pattern matching have proven to be difficult to configure to obtain consistent performance. Particularly when ported from other vertical applications, data loss prevention techniques that attempt to recognize person names, street addresses, phone numbers, email and IP addresses, identifying numbers such as Social Security Numbers, dates and times, etc. have difficulty when confronted with valid strings that may occur in a medical context. Partial rather than complete matching and redaction has also been observed [Kopchick 2022].

We reiterate that it is irrelevant to the effectiveness of de-identification whether offending text occurs within structured data elements identified as being at risk, or elsewhere in the collection, such as in unstructured text. OCR guidance on the HIPAA Privacy Rule specifically calls attention to this: *"The de-identification standard makes no distinction between data entered into standardized fields and information entered as free text (i.e., structured and unstructured text) -- an identifier listed in the Safe Harbor standard must be removed regardless of its location in a record if it is recognizable as an identifier"* [OCRd].

The statistical implications of re-identification risk posed by unstructured text has been considered [Batet 2014].

### 1.17.3 Private Data Elements and Concepts

The matter of private data elements deserves further attention. These are data elements that the DICOM standard allows to be included, but whose meaning and encoding are not defined by the standard itself. A self-registration scheme allows each implementer to define their own set of private data elements. Their meaning is supposed to be published in the vendor's DICOM conformance statement. Though removal of these private data elements does not formally affect conformance of the image with the standard, it may sacrifice information of considerable utility for some applications. As technology evolves, and before features have been added to the DICOM standard, vendors may document important values in a structured form in private data elements, which might otherwise be available only in unstructured text description fields. For example, helical scan pitch and iterative reconstruction methods in CT, or diffusion B-value in MR, are examples of technological improvements whose widespread deployment in the field predated their incorporation into the DICOM standard.

Retention of such private information, when safe, greatly enhances the utility and reuse potential of the data. In addition, it is a sad reality that the full functioning of some vendor-supplied advanced quantitative analytic and processing tools depends on a complete set of their expected private data elements. So, their retention may be essential for completion of the task. DICOM Supplement 142 [DSC Sup 142] introduced the concept of private data elements that are "known safe" (with respect to privacy disclosure) and began to tabulate them in the standard. The formally published list in the standard [NEMA PS3.15 Safe Private] remains relatively short but is actively maintained and extended. Other third parties have made their own lists, based on their own assessments with respect to safety. The most well-known of these that has been shared publicly is from The Cancer Image Archive (TCIA) [TCIA De-id

Overview]. Subject matter experts may be required to determine which private data elements need to be retained to preserve utility [Kondylakis 2024].

When processing private data elements, care should be taken to preserve the private creator, as well as to consider that the reserved blocks to which they are assigned are relocatable.

Some vendors are known to mix text and binary components within their own proprietary structures in relatively large opaque binary values buried in private data elements of undisclosed format. These may need to be discarded unless known to be safe but may also be critical for preservation of downstream functionality in some cases.

Analogous to private data elements, and requiring similar treatment, are coded concepts. These may be used as the values for traditional data elements (such as Procedure Code Sequence), or as the names or values of coded name-value pairs in structured content such as structured reports (whether they be encoded in DICOM SRs or in some other format, like HL7 Clinical Document Architecture (CDA) [Dolin 2006]. These are further discussed under the subjects of profiles for rule-based de-identification and reports, documents, and annotations. The bottom line is that if the coded concept is not known to be safe or cannot be assumed to be safe based on some heuristic, then it may need to be removed or replaced. This is particularly true for SR content items that are of value type TEXT, i.e., have an (unknown) coded name and an unconstrained text string value. Note that even though some coded concepts may not be included in the DICOM standard, they may well be defined by some reliable standard lexicon or ontology (such as SNOMED CT), and there may be opportunities to divine their safety from such sources. E.g., all coded concepts that are sub-classes of the SNOMED CT concept for "anatomical structure" might be deemed safe, even if the context in which the concept is used as a value is unrecognized.

Private data elements in a specified block of elements are identified by a string private creator value encoded in another data element, which allows them to be relocated to different blocks. De-identifying tools need to be insensitive to which block is used and recognize the creator, and this needs to be tested. Private creator values themselves will not normally contain PII and are safe, as long as a restricted list of known safe private data elements is used, the creators of which should already have been proactively vetted. However, private creator values will typically be attributable to a particular manufacturer and model of device. Also, if there are no known safe private data elements being retained for a particular private creator, then that private creator is not needed and should not be retained either.

### 1.17.4 Value Representation

DICOM data elements are relatively strongly typed, in that there is a Value Representation (VR) specified for every data element. The VR is either encoded explicitly in band or known from the data dictionary. The VR can be used to provide guidance for handling an unrecognized data element that otherwise might have to be removed. For example, if it is a date or a floating point number, then a generic handling decision might be applicable rather than removing the value entirely; conversely, if the VR indicates that the value is unstructured text, then removal or text cleaning might be initiated. Recognizing date and time data elements may be of use for triggering generic datetime handling [Suzuki 2007], but since related dates and times for the same event are often encoding in separate DA and TM data elements, rather than a combined DT data element, additional knowledge beyond the VR about the relationship may be needed. This is not explicitly encoded in the standard and requires a manual or heuristic approach based on the keyword or name to extract (e.g., determine that Content Date and Content Time are related).

For private data elements, or for new standard data elements, if the VR is not encoded in the dataset (so-called "implicit" rather than "explicit" VR), then the VR is deemed to be unknown. There are special mechanisms in the standard for encoding and re-transmitting Unknown VR (UN VR), but from the perspective of de-identification the same principle applies. If a data element has an unknown VR, and it cannot be determined from a dictionary, then either the opaque binary content of the value needs to be analyzed or the value needs to be removed or replaced. Since such data elements may contain nested sequences, the analysis of such unknown BLOBs may be non-trivial.

The handling of those VRs that contain text, including CS VR, is further discussed under the heading of text in structured data elements.

Even though data elements may be nominally numeric, whether encoded as binary (e.g., US) or string values (e.g., IS) it may sometimes be necessary to consider them for cleaning as text. For example, a date (as digits without delimiters) might theoretically appear in an IS (e.g., Series Number). In general, such data elements are not listed in DICOM PS3.15 as needed removal or replacement.

### 1.17.5 Retention of DICOM Compliance

No matter what approach is chosen, rule-based or statistical, whether text is cleaned or removed or replaced, it is vital to preserve the compliance of the encoded DICOM image with the DICOM standard [Lien 2011]. This includes retaining required data elements and values for compliance with the Information Object Definition (IOD), replacing rather than removing them, if necessary, as well as using valid values and codes. It is important to recognize that some minimal sets (of data elements) suggested for retention (e.g., [Parker 2021b]) are insufficient to maintain compliance of the de-identified object with the standard IOD, which greatly compromises their functionality and utility, despite caveats about it being desirable to retain more information for preservation of scientific utility. Further, not infrequently invalid values are used as replacements, e.g., rather than being removed, if optional or permitted to be empty, or replaced with a dummy date of valid form (e.g., "19700101"), sometimes a date (DA VR) will be replaced with a string like "ANONYMIZED”, creating an invalid DICOM object.

It is essential, in our opinion, that the fidelity with respect to DICOM conformance of the de-identified image object be no worse than that of the image that was originally supplied, recognizing that the input may not always be perfectly compliant. Accordingly, the DICOM de-identification profile defines specific actions for each data element such that conformance will be preserved, either in terms of the value's removal, replacement with a zero-length value or a dummy value. Further, we recommend that automated tools be used to validate this conformance [Clunie dciodvfy] after processing, as is done for example, in the process used during TCIA curation [Bennett 2018a].

### 1.17.6 Format Conversion

It is commonly claimed that converting source images from their original DICOM format to some other format reduces the burden for de-identification:

*"Conversion to NIfTI strips the accompanying metadata from the DICOM images, and essentially removes all Protected Health Information (PHI) from the DICOM headers. Furthermore, skull-stripping mitigates potential facial reconstruction/recognition of the patient"* [Baid 2021].

However, depending on the tool used to do the conversion, some of the DICOM metadata may be propagated into the converted file [Irving 2017], while creating an incorrect impression of complete removal of PII.

Further, this conversion approach frequently leads to significant loss of utility:

"*Totally removing the DICOM metadata for opensource research efforts prevents privacy issues but reduces the value of data, because metadata is important for AI algorithm development"* [Willemink 2020].

This loss of utility applies not only to AI applications, but many other traditional forms of image processing, analysis and quantification.

The recognition of the importance of preserving important metadata has led to the development of new DICOM-like formats that replace or augment traditional research formats like NIfTI, such as BIDS [Gorgolewski 2016]. To the extent that the metadata components of such formats are automatically produced from the original DICOM files, they do need to be subject to the same controls as the corresponding DICOM data elements. In some cases, acquisition metadata is preserved but subject (patient) metadata is not, but the known risks for such things as descriptive data elements remain. Of course, if the subject metadata is removed and not replaced with pseudonymous identifiers, the files are no longer self-identifying, and out of band management systems, databases, file structures, and manual processes are required to track the experiment. Patient descriptive (rather than explicitly identifying) information (e.g., for image quantification) needs to be handled in the same manner as any indirect identifiers, from a statistical re-identification risk perspective, regardless of the format.

Some authors share their converted data relatively freely and make the DICOM files available *"upon an argued request"* [Lorenzini 2022].

### 1.17.7 Pseudonymization

As noted in the scope, the general procedures for, and applicability of, pseudonymization, are out of scope for this report. More detailed descriptions can be found in the literature [A29WP 2014]. We will only address the matter to the extent that some aspects of pseudonymization are relevant to the successful de-identification of image collections with preservation of utility. Systems that separate potential identifying metadata from the images and selectively reassociates it using pseudonyms are out of scope [Abouakil 2011].

To state the obvious, when a collection contains more than one subject, it is useful to distinguish the subjects from each other, as well as to link records about the same subject within the collection [Suzuki 2007]. Linking of de-identified records is distinct from the need for authorized re-identification of subjects [Kohlmayer 2019]. Such linkage is generally achieved by giving the subjects new individual identifiers, which are different from their original identifiers, i.e., pseudonyms, and unique within the collection. The replacement identifiers may have been pre-defined according to some scheme (such as in association with a randomization schedule for clinical trials subjects) or be generated by some random, arbitrary or cryptographic process. Various tools and trusted third party tools have been created for performing pseudonymization on demand [Wündisch 2024].

In practice, the DICOM Patient's ID is usually populated with the pseudonymous identifier, which is also typically propagated in some form into Patient's Name and Clinical Trial Subject ID. In some cases other identifiers may also be replaced, such as Patient's Birth Date [OMI-DB] but in most cases this is emptied, it not being necessary, and the need to ensure any replacement value remains consistent with the patient's age, which may be need to be independently modified itself as an indirect identifier.

When an entire collection is de-identified from submitted data at the same time, it is straightforward to replace all instances of the original identifier with the pseudonymous identifier, no matter whether a persistent map or hashing technique is used.

However, when a collection is being incrementally extended over time, perhaps as new imaging studies have been received, it becomes a challenge to associate the same replacement pseudonym with the same original identity. This is less of an issue when adding subsequent analysis results, since those can be identified using the pseudonyms supplied in the images being analyzed.

Issues associated with the collection of sequential events and the role of pseudonymization have been explored specifically in the context of medical imaging [Noumeir 2007] [Patel 2016]. A mapping rather than hashing approach obviously requires persistence of the map and moves the process into the realm of "reversible pseudonymization" in that someone potentially has access to that map. If the hashing mechanism is parameterized (e.g., with a key or salt), then to be reproducible, the parameters need to be persisted and reused. Note that if dates are preserved in the images, either exactly or in a modified form retaining the integrity of their temporal relationship, this also needs to be accounted for, as discussed elsewhere.

If hashing approaches are used, considerable care needs to be directed to the choice and implementation of the mechanism [AEPD 2019] [Finck 2019].

In the real world, pseudonymization can be significantly complicated by the fact that patients are sometimes not easy to identify reliably in the first place, as their names change (e.g., with marriage) or they visit different facilities (especially in the absence of unique national identifiers). Such issues are largely out of scope of this discussion and are mentioned only to the extent that they impinge upon the difficulty of tracking longitudinal information about individuals such that it is reliably and accurately represented in publicly accessible collections. We note that both the mapping and hashing techniques, but particularly the latter, are vulnerable to small variations in the form of the input string that is the original identifier. Hash-based mechanisms designed to match people based on multiple keys rather than a single identifier have been devised [Churches 2003], but are out of scope for this paper.

In practice, many large public collections are not released until well after the acquisition of images has been completed (e.g., after some terminal event such as death, progression of disease, or going off study for some other reason). Issues with identity reconciliation are generally resolved by curation prior to de-identification. In passing, we note that though legal requirements for de-identification and pseudonymization may be less rigorous for deceased individuals in some jurisdictions (e.g., being outside the scope of GDPR), we assume from an ethical and moral perspective that the same technical requirements and considerations are applicable.

There may be several cycles of pseudonymization, or several alternative pseudonyms for each subject. When subjects are considered for eligibility for a clinical trial, they may be assigned eligibility pseudonyms. When enrolled they may be assigned on-study identifiers. For various reasons, these clinical trial identifiers might not be those intended for public release, and so new pseudonyms are assigned for that purpose. When data from multiple sites is centralized, the pseudonyms used by the submitter may not be those decided upon by the coordinating center, if for no other reason than their scope of uniqueness may not be sufficient (they may collide with those from other submitters). Yet again, new pseudonyms may be assigned. This may be challenging when different types of data are submitted to different sharing entities, who then need to coordinate record linkage to the same subjects [Dibben 2015]. E.g., imaging, genomic, and proteomic data for the same subjects may be publicly shared by different groups. Coordination of pseudonyms (and indeed re-identification risk assessment) across groups may be needed.

### 1.17.8 Age - Including Pediatric and Geriatric

As has been mentioned for PET SUV, special considerations may apply with respect to preservation of patient characteristics that may be indirect identifiers posing re-identification risk. Children are not just small adults [Larcher 2015]. The spectrum of disease encountered in children, including malignancies, is different. Children are unable to give consent for data sharing, which though it does not affect technical de-identification requirements, may be a factor in determining whether public release is feasible in the first place, and ethical or moral even if nominally legal.

The most obvious aspect for release of individual record data containing patients of very advanced age is their reduced number in the general population, and hence the increased risk of re-identification. This is reflected, for example, in the specific treatment of an age of >89 years in the HIPAA PRSH 18 elements [OCRb]. If a formal statistical analysis of re-identification risk were to be performed, it would include this consideration in a general manner, but in the absence of such, if age is to be included in the public data release, some consideration should be given to this factor. Since the distribution of ages in the collection (sample) is likely very different from the distribution of ages in the population at large, this factor should be considered in the estimation of population uniqueness and its impact on re-identification risk.

The expert determination method may theoretically allow greater preservation of information than the rule-based HIPAA PRSH method. For example, more precise information about elderly patients >89 years might be retainable. This may become particularly relevant as the population ages dramatically [Bloom 2016] and the elderly become less unique. It also has implications for imaging studies of the elderly, particularly those in which chronological age is a covariate.

The utility of age for a range of qualitative and quantitative secondary use cases needs to be considered, as well as the impact of any quantization or other generalization.

Consideration should also be given to the unfortunate manner in which age is encoded in DICOM objects. Specifically, it is a positive integer string with exactly three numeric digits and units of days, weeks, months, or years. It is not possible, for example, to explicitly convey in de-identified DICOM objects that age has been quantized into bins of, say, decades, and a suitable single representative replacement value needs (e.g., "090Y" in the case of DICOM) to be selected and documented. While it is certainly the case that additional age information can be conveyed out of band (e.g., in an associated clinical data spreadsheet), or in private data elements or even new standard data elements, that would defeat the goal of retaining the utility of existing standard data elements for use in standard software. Care needs to be taken when combining collections that have previously been de-identified to assure that the same replacement value for generalized ages (e.g., top coded > 89 years) is used in the combined collection.

The choice of age quantization strategy might be very different for neonates as opposed to older children, adolescents, adults, and geriatrics. This raises the question of what to do for a collection that is comprised of a broad spectrum of ages, which would entail a non-uniform approach that minimized re-identification risk while preserving maximum utility.

Patient's age in the standard DICOM data element is the age at the time of acquisition of the image, as specified by the device operator or information system that provided the modality work list. As such it may not be entirely accurate. There may be other age-related events that are of clinical relevance to understanding the images in the context of a set of time points, such as age at diagnosis. This information is typically not included in the image data elements, but in associated clinical data spreadsheets or databases. Ideally, a consistent approach should be used when removing, blurring, or adding noise to all ages associated with a specific patient across all the images for all the time points for that patient.

The extraction of age from image features even when not explicitly encoded is discussed elsewhere.

### 1.17.9 Dates and Times

The presence of actual dates and times in image metadata and accompanying clinical data may pose a re-identification risk, depending on what real-world event they are attributed to, and what other information an intruder might have access to.

Dates and times have specific risk profiles depending on the event to which they are attributed, and they may have significant implications for successful understanding (and hence re-use) of the images. The time interval between images may affect the understanding of the clinical progress of a disease on a macroscopic scale, measured in days, weeks, months or years. On a more granular scale, the analysis of the movement of a contrast agent through the body may depend on a more precise, shorter, time interval, down to fractions of a second. Furthermore, image acquisition periods may span midnight, so consideration of dates and times that are related as a single combined date-time may be necessary.

The HIPAA PRSH calls attention to dates that are *"related to an individual"* [OCRb] and specifically lists birth date, admission date, discharge date, and death date as being of concern. Arguably, the date on which an imaging study was performed can be related to an individual and should be replaced. Certainly, this has led to issues with the release of some image data [MacMillan 2019]. The HIPAA PRSH permits dates with a granularity of year. The same clause calls attention to ages 89 and above, which otherwise must be consolidated into a single category of age 90 or older. These decisions made when the rule was written with respect to the risk of re-identification may or may not be appropriate for the data being de-identified and should be re-evaluated during risk analysis [Benitez 2010] [OCRc].

From the perspective of preserving utility of de-identified data, both dates and times may be crucial, so removal or replacement with meaningless values may not suffice. It may be important to preserve the temporal relationship between images obtained on different occasions, as well as the relationship to clinically or physiologically significant events such as the administration of contrast or a radiopharmaceutical, and other information in the accompanying clinical data. These may affect the qualitative or quantitative assessment of progression of disease, response to therapy, or physiological processes.

The DICOM standard for de-identification specifies that by default all dates and times be removed, but includes named options to Retain Longitudinal Temporal Information, either with full dates, for use when there is no concern about re-identification using dates, or with modified dates. The modified dates approach is commonly used, and though the standard is not specific about the method used, it requires that:
*"modification of the dates and times shall be performed in a manner that:*

- *aggregates or transforms dates so as to reduce the possibility of matching for re-identification*
- *preserves the gross longitudinal temporal relationships between images obtained on different dates to the extent necessary for the application*
- *preserves the fine temporal relationships between images and real-world events to the extent necessary for analysis of the images for the application"*

The DICOM description includes both dates and times in this description, since it is presumed that conceptually, even though they may be encoded in different data elements (like Study Date and Study Time), the values may be considered together as comprising a date-time stamp with high precision that is independent of the time of day. The combined date-time information also supports time periods that span midnight, such as a dynamic contrast study that might begin just before midnight on one day and end just after midnight on the next day. Without considering the date and time together, the duration of such an acquisition and the dynamic contrast relationship of successive slices would be difficult to determine.

From a de-identification perspective, modification of dates and times may consider date alone, dates and times independently from each other, or dates and times in combination with each other. TCIA, for example, has determined that it is sufficient to modify dates only, and leave times alone (including dates within combined date time data elements) [TCIA De-id Ovr]. Other tools have been described that modify dates to a random epoch consistently for a specific subject over successive de-identification sessions [Kundu 2020a] [Kundu 2020b]. Some tools modify the dates and times together as if they were a single timestamp [Clunie DicomCleaner].

Shifting dates [Hrynaszkiewicz 2010] and times in correctly formatted structured date elements is straightforward and should be safe. Dates and times may also be present in unstructured text or burned into images. If burned in, they should probably be removed rather than replaced with a shifted value. If dates are recognized within unstructured text and there is an effort to shift rather than remove them, it is possible to inadvertently disclose the amount dates are being shifted, if the recognized text is not actually a date after all, but something else whose original value is predictable from context. This may result in the original values of all dates that were shifted by the same amount being recoverable [Alexander 2022].

Other considerations in choosing a date-time shifting method include the presence of partial dates, and the potential need to preserve some temporal proximity to significant real-world events [Reppert 2018]. If the shift is randomized but may be forward or backward, consider that zero might be included in the random interval and no shift would occur.

### 1.17.10 UIDs

DICOM makes extensive use of ISO OSI-style unique identifiers (UIDs) [NEMA PS3.5 Annex B] for the purpose of identifying standard classes of objects as well as specific instances of entities. These are referred to as OIDs (object identifiers) in HL7 standards [Steindel 2010].

For example, the class of an information object is specified by its standard SOP Class UID, which is one of a fixed set of values defined by the DICOM standard, or of private values defined by an implementer. Such non-transient UIDs need to be preserved as-is in any de-identified image, otherwise the information object and its structure and encoding will not be recognizable.

Transient UIDs are created by the implementer to represent instances of entities, such as studies, series, and SOP instances (including individual images). Though no UID is usually assigned to the patient per se, these transient UIDs may be considered sufficiently specific to the patient as to constitute direct identifiers; i.e., a SOP Instance UID of a specific image of a patient belongs to no other natural person than that patient. Even though matching of such transient UIDs related to DICOM objects would require an intruder to have access to a registry of such information anyway, in which case they would arguably have the images and their metadata as well, it is relatively straightforward to replace all such UIDs. Also, the UID generation process may result in embedding of date/time, device or site identifying information. The DICOM standard requires that SOP Instance UIDs be changed only when creating derived images if changes to the pixel data affects professional interpretation [NEMA PS3.3 C.7.6.1.1.2]. Since de-identified images will not normally co-exist in the same system as their original predecessors, it might not be absolutely necessary to change UIDs, but it is generally considered good practice.

However, such UIDs cannot just be replaced randomly, since they are used to both group instances of the same entity (e.g., images in the same series must have the same Series Instance UID) as well as to refer to other instances in other objects (e.g., from a transverse image to the localizer image on which it was prescribed, or to the source image that was processed to form a derived image, or the image or 3D frame of reference to which an annotation applies). In other words, within a defined scope of referential integrity, UIDs must be replaced consistently [Noumeir 2007] [Prior 2020]. That scope must at least include all objects within a single DICOM study, and preferably all objects for a single patient that may have multiple studies.

One of two techniques is generally used to maintain referential integrity:

- a persistent map (e.g., database) of original to replacement UID, used such that whenever a new UID is encountered, a replacement is created and used and entered into the map, and reused the next time that UID is encountered, or
- a hash-based mechanism that generates a replacement UID from the original UID by a one-way cryptographic function that is impractical to reverse or attack in a brute force manner, and deterministic in that it will produce the same replacement each time it is requested; a salt may be used to reduce vulnerability to brute force dictionary attacks [Lien 2011].

The map technique is robust but depends upon the map being persisted for as long as it needs to be used, which is generally not beyond a single session of de-identification activity. This means that if new objects are added to a previously de-identified study after the map has been discarded, they will not be assigned the same UIDs. Further, if the de-identification session is parallelized across instances in a study, the map must be shared between the processes or threads engaged. While it might seem desirable to preserve such a map in the long term, for reuse, error correction or audit trail purposes, if this is done, it must be securely protected.

The hash technique seems simpler but runs the theoretical risk of all cryptograph techniques, and that is being eventually "broken" by analytic or computational techniques [Sharma 2018]. Publicly shared image collections will potentially live forever, so at some time in the future, the possibility exists that a hashed value could be reversed. Though, as noted earlier, most intruders will not have access to the original UIDs to match against anyway, so there would be little if any justification for the effort involved.

The hash technique is one of the options provided by the well-known RSNA MIRC Clinical Trial Processor (CTP) tool [RSNA MIRC CTP]. The approach is to append an MD5 hash of the original UID to a root prefix and truncate it if it exceeds the maximum permitted length. See the source code [CTP Source] of java.org.rsna.ctp.stdstages.anonymizer.AnonymizerFunctions.hashUID() for further details. MD5 is known to be vulnerable to collision attacks [CERT 2008] but that does not necessarily make it unsuitable for this use. Truncation of hash values when the root prefix is not short enough is a recognized disadvantage of this method [Lien 2011].

It goes without saying that the replacement UIDs need to be valid UIDs in that they must be correctly formed, are indeed globally unique, do not risk collision with other UID assigners, and follow all other rules of DICOM UID assignment and encoding. This is not difficult but requires attention to detail. We mention this because some de-identification tools have been observed to generate invalid UIDs.

The HIPAA PRSH makes references to *"any other unique identifying number, characteristic, or code"* [OCRb], but it is generally interpreted to refer to unique identifiers of the patient per se, and as noted, DICOM does not (usually) specify a UID for the patient. However, to avoid the risk of an overzealous interpretation of this provision, replacement of all transient UIDs may be prudent.

The DICOM baseline de-identification profile [NEMA PS3.15 E.1] requires all transient IODs to be replaced but does provide an option to retain UIDs if they are needed for some purpose (the Retain UIDs Option). It is recommended that if UIDs are retained, that the reasons for doing so (utility) and the consequences (risk) are documented in the risk assessment. The DICOM profile further addresses the processing of UIDs nested in sequences such as Source Image Sequence and Referenced Image Sequence; ideally these should be recursively processed and nested UIDs replaced to maintain referential integrity, rather than completely removed (which would reduce utility and might compromise compliance with the IOD).

Though DICOM does not use them (directly), many other standards and image formats make use of GUIDS or UUIDs for various transient entity identification purposes. The same principles apply to the potential need for replacement of these, and possible mechanisms for doing so.

#### 1.17.11 Original Attributes

Since DICOM provides cryptographic mechanisms for attribute (data element) level confidentiality, it is theoretically possible to encrypt the original values of any de-identified data elements and preserve them in the de-identified image. In theory, only someone with access to the decryption key would be able to decrypt the information and reverse the de-identification.

This is generally inadvisable, since it adds little if any utility, especially for publicly released data and secondary re-use purposes, and potentially adds significant risk. Encryption mechanisms are not infallible and not infrequently poorly implemented, keys are mishandled or subject to insider attacks, and cryptographic schemes may be broken with enough effort.

## 1.18 Image Features

Image content includes both non-biometric and biometric identifiers that constitute an identity risk [Garfinkel 2015]. Here we will consider as potential biometric identifiers those physical features of an imaging subject that are exposed within the image pixel data itself, rather than its metadata, the re-identification risk posed, and potential mitigation. The features may be sufficiently unique as to serve as biometric identifiers, essentially direct identifiers, or be attributes of a subject that serve as indirect identifiers (such as age, sex, or race). The nature of the feature as well as the modality of imaging acquisition need to be considered.

### 1.18.1 Matching Comparable Images

It would be surprising if given two sets of images of the same person in different collections, that one would not be able to match them with a reasonable degree of likelihood. Recent work has demonstrated this unremarkable result by demonstrating the ability to match different chest X-rays of the same subject drawn from within a single large public collection using a machine learning approach [Packhäuser 2022], and the mechanisms behind successful matching have been explored [Macpherson 2023]. Image matching of chest X-rays has been proposed for workflow improvement by detecting incorrect identification [Kondo 2003] [Morishita 2005] [Kao 2013] [Ueda 2023]. Similar considerations apply to genetic material, and the utility of a match depends on linkage to additional identifying information [Nietfeld 2007].

The ability to use comparable images to achieve record linkage, and thus reveal that information about a subject in one collection is applicable to the same subject in another collection, is exactly the same as establishing a match by means of a unique combination of indirect identifiers in an SDC setting. Therefore, the re-identification risk needs to be modelled appropriately, in context, accounting for the likelihood that there will be a near-perfect population register of comparable images available to an intruder. Today, public photographic databases include not only public figures, celebrities, and other famous people, but now social networks capture almost everyone, even those indirectly included in an image such as in the background. A specific effort to model the risks of similarity in a WSI-specific context has also recently been described [Holub 2023]. That work further elaborates on scenarios in which image similarity may create risk.

### 1.18.2 External Photographic Images

Traditional clinical photographic images have long been a concern from a de-identification perspective for the obvious reason that a realistic photograph of an individual may be recognizable, especially if it contains the entire face. Medical photographers generating material for publication have long understood the need to address this concern [Eastman Kodak 1972]. Traditional techniques, such as masking the eyes, have practical limitations, especially if other particularly recognizable traits are present [Slue 1989] [Riis 1991] [Smith 1991] [Roberts 2016]. Until recently, the primary concern has been that relatives and friends of an individual could recognize them. It has been proposed that *"the test of anonymity is whether the patient could recognize his or her own image"* [Robinson 2014].The HIPAA PRSH method [OCRa] [OCRb] specifically includes "*full-face photographs*" amongst those items to be removed.

With the widespread deployment of computerized facial recognition technology and the ready availability of images to match against, whether from social media, government sponsored, or commercial sources, the risk of re-identification has increased. Government-sponsored efforts to improve facial recognition technology have long been a priority [Phillips 2002] [Phillips 2005] [Phillips 2007] in addition to the commercial potential for such services [Glaser 2019] [Metz 2021] [Brodkin 2022]. Progressive improvement over time in facial recognition performance has been observed [Lawrence 1997] [Taigman 2014]. Ordinary consumers already have access to facial recognition services and can use them to identify individuals not otherwise known to them [Reilly 2021]. The state of the art has reached the point that the US allows entry into the country on facial identification alone without requiring a physical document [CBP]. It must at this point be assumed that near-perfect recognition is possible for certain use cases. Regardless of voluntary withdrawal [Grenoble 2021], or enforced efforts to restrict such services through claims of abuse of the private personal data on which they are based [UKICO 2022] [Bagwe 2022], the routine deployment of facial recognition by government and nefarious entities, if not the general public, seems inevitable [NASEM 2024].

From the perspective of de-identification of medical photographic images for public release, if the face is present within the image, it needs to be completely removed or obscured. Though there is a body of research into what components of the face are responsible for recognition by humans or machine algorithms [Karczmarek 2017] there is also concern that so-called "external features" (e.g., hair, head outline, neck, and shoulders) have an influence [Slue 1989] [Jarudi 2003] [Frowd 2007] [Kamps 2018]. Accordingly, any head and neck image may be of potential concern. The use of the structure of the ear as a biometric identifier has also been considered [Hoogstrate 2001] [Emeršič 2017] [Ahila Priyadharshini 2021].

One approach to mitigating such concerns is to share only relatively closeup photographs of clinical findings, which excludes the entire face, or head and neck. Other approaches address the modification rather than obscuration of the face to prevent de-identification [Engelstad 2011] [Maximov 2020] [Kuang 2022] [Yang 2022], but it is uncertain how such changes would affect the utility of the images for demonstrating the clinical feature for which the image was being shared, or their potential for secondary reuse.

Though not directly related to the problem of recognizing faces from photographs (or from cross-sectional images that have been reconstructed), there has been research into facial recognition from 3D surfaces from optical surface scanners [Kim 2017].

A completely different approach to recognition is examining the skin detail rather than the facial structure [Pierrard 2007] [Park 2011]. Potentially this could be generalized to photographs of any part of the body.

### 1.18.3 Reconstructed Faces

Faces can be reconstructed with high apparent fidelity, even from several thousand year old specimens [Cesarani 2004], even if recognition is not a consideration. The potential to match such reconstructed images with actual individuals using photographic images available to an intruder has long been a recognized theoretical problem, and one that some in the neuroimaging community have proactively considered [Kulynych 2002] [Toga 2002].

We distinguish the techniques that achieve indirect identification by matching to databases of facial photographs, from those that are intended for forensic matching of internal structures based on 3D matching techniques [Dong 2022], or from those where the reconstruction is from functional MRI signals from an observer's brain [Zhang 2020], which will not be considered further here.

#### *1.18.3.1 Success Rate and Re-identification Risk*

Early experiments with testing the performance of such matching produced modest results with human matchers [Chen 2007] [Prior 2009] and automated tools [Mazura 2012]. The more widespread availability of reconstruction tools coupled with the rise of high performance facial recognition services with large databases has led to an increase in the level of concern, and more recent experiments have suggested an improved rate of success at matching using readily available tools [Parks 2017] [Grother 2018] [Schwarz 2019] [Schwarz 2020] [Schwarz 2022]. [Schwarz 2023a].

One commenter stated: *"Neither study [Schwarz 2019 or Mazura 2012] adequately predicts the performance of face-recognition software in matching a small set of rendered images to large databases containing millions of photographs, as may be attempted by a malicious agent"* [Juluru 2020].

That said, the steadily improving performance has concerning implications, even if the risk of a sample to population threat model (journalist attack) has not been directly quantified. Further, the threat model of identifying a specific sample known to be in a collection (population to sample, prosecutor attack) is not to be dismissed, given the potential for risk to the reputation of the data holders and to re-identified participants.

Small experiments may theoretically be affected by data set bias, since it has been shown in the past that *"high classification accuracy of face recognition datasets can be achieved without recognizing faces"* [Model 2015] [Shamir 2008]. However, this factor is not applicable to the Schwarz or similar experiments, because there is nothing shared between the background of the photographs and the scanned images.

Unfortunately, contemporary articles about data sharing [Wichmann 2020] [Diaz 2021] quote uncritically the 83% MRI-to-photograph matching success rate from [Schwarz 2019] without critically considering its real-world applicability. This is not to say that the matter is not of concern, quite the converse, especially since performance is improving, now 98% for MRI [Schwarz 2022], though recent simulation studies suggest these rates substantially overestimate the real-world re-identification risk [Jwa 2024]. Rather, like all aspects of de-identification that balance risk of re-identification against preservation of utility, the measures used to guide reasonable decisions must be realistic and defensible. The need to achieve a reasonable balance has long been a subject of concern in the neuroimaging community [Bischoff-Grethe 2007] [White 2022], and is increasingly apparent to those gathering images for AI applications [Lotan 2020].

The absence of any known successful re-identification attack using facial recognition, despite the long-standing widespread unconstrained public availability of large numbers of CT and MR head image sets, would suggest that the sky is not falling yet. Overreaction in the absence of robust statistical modeling of the risk has already led to reduced utility, especially for re-use of information beyond the cranial cavity, such as for head and neck cancer. Even within the cranium there are utility issues with images that have been de-faced [de Sitter 2020], though these may be small [Schwarz 2020] [Buimer 2021] [Theyers 2021] [Rubbert 2022] [Bhalerao 2022] [Schwarz 2022] [Schwarz 2023b] particularly when compared to other sources of error, such as rescanning [Gao 2022b].

Since current studies are inadequate with respect to risk [Juluru 2020], and significant swaths of head and neck images have recently been withdrawn from unrestricted public access on these grounds [TCIA 2022], it is difficult for this task group to recommend in this report anything other than caution in the absence of a meaningful risk analysis. Even the ADNI initiative, which had long avoided the use of de-facing, has now added it for future phases [Schwarz 2024]. The doctrine of considering images of the facial region as potentially identifiable is now being included in basic radiology training [ABR 2022].

It is our assertion that any such risk analysis should follow the same principles as the SDC community uses in terms of selection of an appropriate context, threat model, risk model and risk estimation methodology. In other words, reconstructed faces (and indeed actual photographic images of faces), should be treated in a similar manner to indirect identifiers in structured data. The risk of establishing uniqueness in the population as well as successful confirmation of such a match should be considered in a formal statistical manner. Until an appropriate statistical model of estimating population uniqueness of a reconstructed face from information in the image collection sample has been produced, the alternative method used in the SDC community of comparison against a population registry that is as complete as possible may need to suffice. Given that various entrepreneurs have an avowed goal of producing a complete registry of faces [Brodkin 2022], such an estimation may be possible. It should go without saying that the same defensible thresholds of re-identification risk acceptability should then be applied, since the issue is one of re-identification alone, no matter by what means it was achieved (facial recognition as opposed to record linkage by indirect identifiers).

We note that evaluation methods that are well established in the traditional facial recognition community [Phillips 2000] may or may not align with the threat models for SDC. For example, a common use case for facial recognition is a watch list of suspects, which contains *"... elements of both the verification and identification tasks ... In the watch list task, a photo of an unknown person is provided to a system and the system must determine if the unknown person is on a watch list"* [Phillips 2002]. Different approaches are used when the watchlist is small or large [Kamgar-Parsi 2011]. There have been attempts to combine SDC concepts such as k-anonymity with facial recognition prevention [Nakamura 2021], though not yet in the context of reconstructed cross-sectional images.

An additional factor to consider may be how recognizable various individuals are, either because of unusual features, or of being particularly well known (celebrity status) or of being known personally to an intruder. Another consequence of celebrity status is that the probability of images being available for comparison is higher. Sex, race, and age may have an impact on recognizability [O'Toole 2017] [Jwa 2024]. Accordingly, perhaps not all individuals should be treated equally in the risk model. Historical lessons from the experience with de-identification of photographs of faces for publication may be applicable [Slue 1989].

#### *1.18.3.2 Coverage and Sampling*

In addition to the modality [Schwarz 2022], other significant factors in the success of re-identification are:

- the coverage of the scan (both from side to side and top to bottom)
- the thickness and spacing of the slices
- the type of acquisition [Schwarz 2023]

When voxels are not isotropic, the plane of the slices may also be expected to be significant, though that may be partially correlated with the coverage.

The clinical indication for the scan dictates both the coverage and the thickness and spacing, though this may change over time as technology and the standard of care evolve. For example, standard of care MR images for brain tumors are often very different in both coverage and sampling when compared to MR images for volumetry of the whole brain or hippocampus for dementia, tumor imaging being more restricted in coverage and utilizing thicker slices. Standard of care CT brain images may well not include the entire face and be relatively coarsely sampled unless the objective is a 3D reconstruction of the bony anatomy.

With respect to thickness and spacing, a more than 5-fold difference in recognition rate has been demonstrated between 3.5mm and 6mm CT slices [Parks 2017]. For many of the experimental studies, neither the coverage nor the thickness and spacing represent the norm encountered in clinical care, so their matching rates may be misleading. That said, images acquired primarily for research may well use quality greater than routine clinical care.

On the other hand, incomplete coverage may be addressed by attempting to synthesize the missing sections to produce a complete face for matching attempts [Schwarz 2022]. This begs the question of what facial recognition software is actually matching on, but if the attempt is successful then the issue is moot. A more important question may be to determine just how much anatomical coverage is necessary to create a significant risk of matching.

Though it is traditional to consider scans of the brain as being of primary concern, there are other reasons for scanning that may include part if not all of the face. Head and neck studies for carcinoma have already been mentioned. Whole body PET/CT scans typically include the neck and skull base, for example, for the purpose of locating lymph node metastases. Though they may not include the whole brain and skull, they may include the orbits and supraorbital ridges and hence though they are often relatively thick slices, could pose a theoretical re-identification concern, which should be quantified.

To use an SDC analogy again, the bottom line is that the context of the images may well significantly affect the re-identification risk, and should be considered in the estimate.

#### *1.18.3.3 HIPAA Privacy Rule Safe Harbor Comparable Images*

As previously mentioned, there is a provision in the HIPAA PRSH method that *"comparable images"* to full face photographic images need to be removed. It has not been established that projections of 3D reconstructions are indeed comparable, from a technical, statistical, or legal perspective. Further, this provision is only applicable to the PRSH method, and application of the expert determination method might draw a different conclusion based on statistical estimates of re-identification risk.

#### *1.18.3.4 Mitigation — De-facing and Skull-Stripping*

Analogous to the generalization techniques applied to indirect identifiers in SDC re-identification risk reduction, it is possible to process image data that might otherwise allow reconstruction of a potentially recognizable face. The processing techniques fall into two general categories:

- de-facing, which removes or replaces the facial structures (only)
- skull-stripping, which removes everything outside the cranial cavity (including but not limited to the face)

Key to determining the suitability of these techniques is:

- measuring their effectiveness in reducing re-identification risk,
- such that it can be balanced against the preservation of the utility of the images.

At one extreme, such techniques are unsuitable for head and neck cancer images since such tumors involve structures of the face [Sahlsten 2023]. Therefore, if the estimated re-identification risk of the reconstructed faces exceeds an acceptable threshold, then the images cannot be released publicly without restrictions. At the other extreme, a skull-stripped version may be satisfactory for images that involve strictly intracranial processes, and for which secondary uses involving extracranial structures can be sacrificed in the interests of releasing data publicly. The routine use of de-facing is increasingly prevalent for publicly shared data (e.g., for functional neuro-imaging and for aging and dementia imaging), and is preferable to not sharing the data at all.

Though skull-stripped images are *prima facie* without faces, and may be sufficient, we emphasize that skull-stripping is not necessary in the presence of a suitably robust de-facing technique [Silva 2018b]. De-facing may be preferable for data intended for secondary re-use [Schimke 2012] and simulation studies confirm sufficient reduction of the real-world re-identification risk from de-faced images [Jwa 2024]. Even for nominally intracranial processes, such as primary brain tumors, the involvement of the meninges and cerebral venous vasculature may have significance and may be compromised, being at the boundary of what is typically removed by skull stripping or de-facing. The impact of skull-stripping has recently been quantitatively evaluated for meningiomas [LaBella 2024]. It has been demonstrated to negatively affect brain age prediction [Cali 2023]. That said, it may well be possible to find a suitable generic robust de-facing process,  particularly one that is multi-modality (CT, MR and PET) [Schwarz 2022] [Patel 2023], as is also the case for skull-stripping [Hoopes 2022]. Attempts have been made to implement only the minimal changes necessary to protect privacy [Milchenko 2013] [Greve 2022]. The importance of evaluating the impact not only on derived quantitative features, but also on the spatial positioning of remaining structures, as well as the need for human quality control to detect catastrophic failures, has been emphasized [Gao 2023]. Ethnically-specific solutions (such as different templates) may be needed for different populations [Gao 2022a]. In short, in keeping with the philosophy of maximum utility for unanticipated secondary re-use purposes, the minimal distortion possible should be selected to meet the *a priori* established re-identification risk threshold (e.g., de-facing rather than skull-stripping if possible). Even if skull-stripping was used in the pipeline in which the original data was processed (e.g., prior to segmentation of intracranial structures), and those images are already available, it is preferable to also reprocess the original images with de-facing for the purpose of sharing, to maximize their re-use utility.

It is uncertain to what extent more creative artificial-intelligence-based techniques, such as have been proposed for facial photographs [Maximov 2020], might be repurposed for less aggressive modification of facial structures in scans than complete removal. The impact of blurring faces in photographic images on other classification tasks has also been examined [Yang 2022]. One approach for photographs integrates the SDC concept of k-anonymity to attempt to bound the probability of identification [Kuang 2022].

It has been proposed that "masking" of the face by interposing a barrier in 3D space between the face and the potential observer could be used. This would need to be more than trivially reversible, otherwise it would not seem to provide meaningful protection. It has been proposed to have the mask contact the skin surface and be composed of the same distribution of pixel values as the patient data. The intent is to discourage casual attackers and make any re-identification attempt require an unreasonable effort, while preserving greater utility than traditional de-facing approaches [Wardell 2022].

### 1.18.4 Brain Features

Experiments have suggested that brain anatomical features are relatively unique and could be used for identity matching [Wachinger 2014] [Valizadeh 2018]. So too may functional features [Ravindra 2019]. It is presumed that since there is not a widely available population brain anatomic or functional database, which could be used by an intruder for re-identification, this is not a practical consideration for assessment of re-identification risk of publicly released image data [Wachinger 2015], though there may be regulatory considerations in some jurisdictions [Eke 2021]. Obviously, techniques such as de-facing and skull stripping do not mitigate any such theoretical risk [Wachinger 2015].

### 1.18.5 Other Features

Though the face and brain are obvious candidates for re-identification through image recognition, biometrics based on other body parts, particularly the skeleton, have been evaluated in a forensic identification context [Mesejo 2020], such as using CT of the lumbar spine [Decker 2019].

### 1.18.6 Extraction of Patient Characteristics from Images

The role of patient characteristics as potential indirect identifiers leading to increased re-identification risk has been discussed. Here we consider the possibility of using features found in images to determine values for such indirect identifiers, even if they are not explicitly present in the metadata. An important consideration is how accurate derived information would need to be to create sufficiently unique entries. The use of such derived characteristics to estimate a "biological profile" has long been a feature of forensic skeletal identification efforts, and more recently has used both projection radiography and cross-sectional imaging [Mesejo 2020]. Application of AI for non-forensic demographic information recovery from clinical images has also been described [Yi 2021b] [Adleberg 2022].

#### *1.18.6.1 Age*

Even if age is not included as explicit metadata within or accompanying the images, it may be predicted with some degree of accuracy from image features, for some image types and anatomic locations. Though the determination of pediatric bone age is a classic application of plain radiography [Greulich 1959], there are other means of estimating age that may have implications for de-identification.

Mean absolute error is typically around 4 or 5 years for estimates from adult brain MR [Sajedi 2019] [Franke 2019] [Beheshti 2022] though this may improve in the future [Gómez-Ramírez 2022]. With respect to re-identification risk, this might be compared to the amount of generalization of age that is typical when statistical approaches to microdata disclosure control are applied locally or globally. Estimates for younger subjects are more precise [Franke 2012]. A similar or better accuracy for age prediction from chest X-Rays has been observed [Karargyris 2019] [Sabottke 2020] [Yang 2021]. Human observers perform poorly at this task [Chibane 2022].

Also consider that in the absence of an expert determination, the HIPAA PRSH requires ages over 89 to be generalized, yet an approximate age might still be recoverable from the images themselves.

A corollary of this discussion is that further studies on the accurate prediction of chronological age from image features require the presence of precise chronological age metadata, even if the latter is an indirect identifier. This may prove difficult with publicly shared datasets, regardless of whether a statistical (expert determination) or rule-based (HIPAA PRSH) approach is applied.

It is understood that age estimates from images, particular from images of the brain, may deviate significantly from chronological age, particularly in the elderly or in the presence of disease, and the difference itself has been investigated as a predictor of abnormality [Cole 2018] [Franke 2019] [Mishra 2021] [Wagen 2022].

#### *1.18.6.2 Race*

Race may be a significant covariate in many diseases, despite lacking a biological basis [Deyrup 2022]. Even if it is not frequently present in the DICOM header, it is often recorded in accompanying clinical metadata tables. It is certainly a potentially significant indirect identifier and may well be available in registries available to an intruder.

There has recently been research into the recovery of race information from image pixel data testing a range of different modalities [Gichoya 2022] [Burns 2023], though the implications of doing so remain uncertain [Zou 2023]. Whether or not the level of accuracy will rise to the point that it provides sufficient information to assist with a re-identification attack remains to be determined.

#### *1.18.6.3 Sex*

Defining what "sex" is and how it differs from gender [Becker 2022] may be a significant issue when considering what data an intruder might match against indirect identifiers, which needs to be accounted for when evaluating image-derived sex information.

The anatomical sex of a subject can be determined when the gonads, external genitalia, or secondary sexual characteristics are present within the image field of view. Some uncertainty may be introduced due to medical or surgical gender transforming treatments. Indications of such treatment may in and of itself constitute a sensitive attribute disclosure risk, even if it does not contribute to re-identification risk.

It may also be possible to determine sex from other types of images.

Prediction of sex from structural or functional brain imaging may or may not be reliable, may be significantly affected by covariates like brain volume, and may not generalize well [Joel 2015] [Rosenblatt 2016] [Joel 2016] [Sanchis-Segura 2019] [Xin 2019] [Ebel 2022] [Sanchis-Segura 2022]. Prediction of sex from forensic X-Rays has long been described [McCormick 1985], and more recently prediction from clinical chest X-Rays has been demonstrated [Xue 2018] [Yang 2021]. Reasonable machine accuracy has also been found using hand and wrist X-Rays [Yune 2019]. Gender can be predicted from photographic images of faces [Buolamwini 2018] though presumably these would be de-identified as discussed elsewhere in this report.

## 1.19 Embedded Metadata

Identifying information can occur embedded in any part of an encoded image file, and care should be taken to consider all potential locations.

The DICOM data elements have been discussed at length.

The image pixel data payload may also contain identifying information, either within the pixels themselves, as so-called "burned-in text", or in the compression-scheme-specific metadata that is the content of the DICOM Pixel Data element. Indeed, the same is true of non-DICOM pixel data, as has been discussed under image format considerations.

### 1.19.1 Burned-in Text

Distinct from patient "features" implicit in the pixel data content, is text that is present in the pixel data [Robinson 2014]. Some types of images are more prone to the presence of burned in text than others [RCR 2019] [Macdonald 2024]. In most cases this is printed into the pixels in block rather than cursive type (albeit in varying fonts and sizes) and is aligned with the horizontal or vertical axis of the image, and therefore is readily detectable and recognizable using conventional image processing or machine learning based Optical Character Recognition (OCR) techniques [Wang 1997] [Wang 2001] [Chen 2004] [Florea 2005] [Zhu 2010] [Newhauser 2014] [Monteiro 2015] [Silva 2018] [Khosravi 2023] [Macdonald 2024]. Text in a small font size, though it may be human readable, may be inaccurately recognized by machine (OCR) [Osagie 2024]. More recently, cloud providers have also started to assemble complete solutions that combine burned-in text detection and removal with or without metadata processing [Google De-id], [Wiggins 2019] [Microsoft Presidio], though cloud-based services potentially incur performance issues and costs for computation and network transfer, as well as pose security and trust issues. There are other applications for detecting and extracting text burned into images that could be repurposed [Reul 2016] [Vcelak 2019] [Pure Image].

Though the presence of burned-in text is discouraged, and unusual for many modalities, for some, such as ultrasound, it is very common, given a prolonged legacy of video capture rather than direct digital acquisition. Another common reason text is present is when the image payload is actually textual rather than being of the patient. Typical examples are radiation dose screens or tables of measurements. Burned-in text containing identifiers is also usually present in screenshots, whether they be of originally acquired images or derived from post-processing, and their capture user triggered or automated.

Prior to the deployment of general-purpose text detection and recognition techniques that can find text in any location, regardless of size, orientation, font, or content, managing burned-in text often relied on either:

- discarding such images entirely (e.g., anything that the metadata indicated was a "screenshot"), or
- matching the image type, description, vendor and model, and matrix size to pre-defined "templates" of areas where identifiers needing redaction occur.

General purpose OCR tools on the other hand, may be overzealous and potentially remove text that would better be retained, such as acquisition technique descriptors, orientation and laterality information, and clinically relevant but safe annotations and measurements. More precise classification of text content mitigates this issue. Automated OCR techniques can be combined with text classification and analysis, as well as matching with the content of structured metadata such as Patient's Name, to determine what to selectively retain or remove [Vcelak 2019]. The same techniques used for analysis of unstructured text fields, or text occurring in structured but relatively unconstrained descriptive attributes, or indeed clinical documents or radiological reports, can be reused, perhaps augmented to account for the different context.

Though not often encountered, the presence of cursive or hand-written text, or in a particularly small font, particularly in non-orthogonal orientations, can be challenging for some detection and recognition algorithms.

Subtly different, but needing similar treatment, are foreign objects visible in the image that were not burned in per se, but which were physically present. These include lead markers used by technologists during acquisition to signal information such as projection and laterality and other aspects of technique [Khosravi 2023], medical devices with identifying information such as pacemaker serial numbers, and foreign objects such as jewelry, which may contain names or identifiers [MacMillan 2019]. The considerable variation in possibilities of form and content pose a challenge for automated algorithms now and in the future. It is expected that machine learning based algorithms may be retrained to cope as new instances are encountered, but this highlights the need for human review, either complete or on a sampled basis, and constant vigilance for failures.

The need for human review for quality control of burned-in text detection and redaction efforts is revisited later under the subject of Quality Control.

### 1.19.2 Metadata Lurking in Obscure Places

As was mentioned when describing consumer image file formats, compressed data streams may contain their own metadata, and this remains true when the same compression schemes are used either as DICOM Encapsulated Transfer Syntaxes for encoding the Pixel Data, or to encapsulate entire objects such as documents (as described later).

The JPEG family of compression schemes, for example, often used in either lossless or lossy forms as a means of encoding DICOM images, is defined using a series of so-called "marker segments", some of which may contain identifiers or descriptive text. The EXIF standard [CIPA 2012] defines an extensive set of such data elements that are encoded using a TIFF-like syntax, and if the APP1 marker segment that contains the EXIF data is not completely removed, its contents need to be selectively examined and redacted. More recently there have been efforts to support inclusion of more extensive metadata by means of the JPEG Universal Metadata Box Format (JUMBF) APP 11 marker segment [Temmermans 2017] including provenance-related metadata [Rosenthol 2020] and even embedded files, images and external links [DMAG-UPC 2022]. Other standard and proprietary APPn marker segments may be present, and if not recognized as safe, need to be removed. A JPEG bitstream can be parsed and redacted in this manner (removing selected segments) without any need to decompress and recompress the image component payload, so this processing can be performed without loss of image fidelity. The well-known freely available open source exiftool [Harvey] command-line utility is commonly used to examine and edit JPEG metadata and is not limited to only EXIF content. Later JPEG standards have defined alternative syntaxes and means of encoding metadata within the compressed bitstream.

As described later for WSI, dual-personality DICOM-TIFF files may contain metadata in the TIFF header, which is buried inside the DICOM preamble and Dataset Trailing Padding element. This usage is not confined to WSI, however, and was originally developed for ultrasound usage. Accordingly, when redacting DICOM images, care should be taken to replace the 128-byte preamble at the start of each DICOM file, and to remove any Dataset Trailing Padding. If a dual-personality file is to be written, the non-DICOM side of the file should be completely regenerated, just as it is good practice to regenerate rather than attempt to redact the Group 0x0002 File Meta Information. The DICOM preamble has been cited as a potential security risk, which could theoretically be abused by embedding executable code, so every opportunity should be taken to discard its contents when it does not have a defined purpose [CVE-2019-11687].

## 1.20 Modality-Specific Considerations

Though one of the primary benefits of encoding images using DICOM is that most of the content of the header metadata is the same, regardless of modality, images from some modalities have specific peculiarities with respect to the metadata or pixel data that impinge on de-identification.

Some modalities can encode acquired data in a raw form before slices are reconstructed (e.g., CT projection views, MR k-space data, PET list mode or sinogram data). Since there are no formally adopted standards for raw data and it is not often recorded or publicly shared [Chen 2016], it is considered out of the scope of this report. If raw data is required, de-identification is challenging since the encoding is usually proprietary [Langer 2011]. There have been academic efforts to harmonize the encoding of such raw data [Chen 2015] [Inati 2017] [Kesner 2016], and if and when these efforts mature, this issue may need to be reconsidered. If proprietary raw data is encoded in the DICOM Raw Data IOD, then standard data elements can be handled in the conventional manner and private data elements encoding the raw data payload and its description retained only if known to be safe.

### 1.20.1 Ultrasound (US)

Ultrasound images frequently, if not always, contain burned-in identification as previously described [Tessler 2011]. Usually this occurs in pre-specified locations, and if a sufficiently broad range of templates triggered by image size (rows and columns), manufacturer, and model name is defined, can be addressed using relatively mechanical techniques [Antunes 2011] [CTP Pixel], though constant vigilance and maintenance is required as new patterns are created by vendors and encountered.

Even if more modern fully automated general text recognition applications are used, it is important to consider whether burned-in measurement information should be detected and preserved. In some scenarios, these serve as useful annotations, in others a potential source of bias for a subsequent reader [Monteiro 2017].

Though many ultrasound images are single frame, some consist of multi-frame video-like acquisitions, such as for cardiac cine loops. These are frequently lossy compressed because of their size, and during de-identification should not be decompressed (else they become too big) or recompressed (else the loss of fidelity may affect qualitative or quantitative interpretation and fail to meet regulatory requirements for source data). The most common lossy compression scheme used is ordinary JPEG, which operates on 8x8 blocks, so it is possible to redact only the selected zones containing text in compressed space and leave the other blocks untouched [Clunie 2015]. This capability is implemented in various open-source tools that allow manual [Clunie DicomCleaner] or template-driven [CTP Pixel] burned-in text redaction.

Private data elements have already been discussed as an issue in general terms but can be particularly problematic for advanced ultrasound applications. For both advanced quantitative cardiac analysis as well as even basic visualization of 3D acquisitions, vendors may rely heavily on large, opaque, binary, private data elements, the safety of which may be difficult to ascertain. It is important in such cases to communicate with the vendor or review their documentation (especially their DICOM Conformance Statement) to formally determine that it is safe to retain these. Tools, scripts, and templates then need to be adapted accordingly. Ideally, all such known safe private data elements would be documented in the standard [NEMA PS3.15 Safe Private].

### 1.20.2 Positron Emission Tomography (PET)

PET attenuation compensated reconstructed images typically have stored or rescaled pixel values defined as a quantity of radionuclide activity concentration by volume, e.g., with units of Becquerels/milliliter (BQML); the acquisition device is calibrated to measure this physical property, which is not dependent on knowledge of the patient's physiology. Functional evaluation of the information in quantitative terms that are meaningful for evaluating the presence or progression of disease may require transformation based on physiological parameters. Specifically required are the amount of radiopharmaceutical present, its decay during the acquisition, and its volume of distribution. Various metrics, such as Standard(ized) Uptake Volume (SUV), have been defined, which rely on parameterized models to estimate the volume of distribution related to the patient's size [Allen-Auerbach 2009]. Patient-specific parameters required include body weight, height and/or age. SUV based on body weight (SUVbw) is most often used (and has units of grams per milliliter (GML)) but other models have their advantages. Region of interest based calculations can then be further derived (e.g., metabolic tumor volume (MTV)). These concerns are typically applicable to metabolic (18FDG) PET images, but similar computations may be useful for other radiopharmaceuticals (e.g., 18FLT). In oncology, qualitative and quantitative therapeutic response criteria are often defined based on SUV [Kinahan 2015] [Kinahan 2020].

The images that are used clinically and shared for research are typically those produced by the acquisition device. SUV computations are usually made in the display software using the subject-specific size information recorded in the image header or entered manually into the application. The user then has a choice of which model to use, assuming sufficient body size data is available and the necessary header data elements are present to recognize the type of image and perform the calculation [QIBA SUV].

As has been described, patient characteristics may be indirect identifiers or sensitive attributes. Weight, height, and age, particularly if they have extreme values, may contribute to re-identification risk. Whether or not this is the case may be determined by a risk analysis. If the conclusion is that some or all of these values need to be removed or modified, then the utility of a collection containing only activity concentration images would be significantly degraded, since physiological (SUV) images could no longer be computed. In such cases, it may be useful to include in the collection images whose stored or rescaled pixel values have been pre-computed using a specified SUV model. Note that the choice of model will not satisfy every use case. Further, there is a risk that if both the AC and SUV images are included, then the patient-specific model parameters might be back computed anyway (e.g., body weight recovered from a single parameter SUVbw model), defeating the purpose.

If dates and times are being shifted or selectively removed or replaced (see Dates and Times), careful consideration needs to be given to preserving the correct intervals between the scan start time if decay corrected (Series Time) and the injection time (Radiopharmaceutical Start Time) to preserve the quantitative accuracy [QIBA SUV]. Depending on the vendor, some dates and times in private data elements may also need to be shifted by the same amount.

PET studies involving pediatric patients may require special consideration, given that the factors in choosing an appropriate SUV model may differ from adults [Yeung 2002], and the re-identification risk associated with indirect identifiers may be different.

Some PET device vendors do not encode activity concentration images, but instead record the necessary information to compute SUV in private data elements [QIBA SUV Pseudocode] [Lien 2011]. Accordingly, to preserve their utility, it is necessary to retain at least those known safe private data elements during de-identification. As noted earlier though, the patient's body weight might be back computed from the image data in such circumstances.

Since PET/CT images typically require some processing to window, pseudo-color, superimpose and project them in 3D [Im 2010], to support less capable viewers, additional series are sometimes recorded that contain pre-rendered images. These may contain burned-in text identification, so may need to be detected and either excluded or processed to remove the offending text.

The potential for the superior slices of a whole-body PET/CT study to contain information sufficient to reconstruct a face has been discussed. The PET images themselves are not immune to facial recognition even in the absence of corresponding CT images [Schwarz 2022].

### 1.20.3 Nuclear Medicine (NM)

Nuclear medicine images may be planar projections or reconstructed into cross-sectional slices (SPECT). Historically, single projection images, as well as derived images such as 3D reconstructions of SPECT slices and functional model output have often been (or are designed to mimic) screen captures, not unlike ultrasound, and contain burned-in identification in predetermined locations.

### 1.20.4 Radiotherapy (RT)

The radiotherapy workflow generates many images and image-like or image-related artifacts. The development of RT applications using AI are heavily dependent on publicly shared datasets and de-identification issues need to be addressed [Wahid 2022]. For many years, the RT industry has, like diagnostic radiology, adopted the DICOM standard as the universal format for encoding [Law 2009]. Accordingly, there are very few RT-specific de-identification issues and generic DICOM de-identification approaches suffice [Newhauser 2014].

RT objects do contain many data elements that are not used by other specialties, but since these are included in the DICOM de-identification standard [NEMA PS3.15 E.1], modern tools should already handle them appropriately. Descriptive text fields, especially for annotations in RT Structure Sets, are critical to the understanding of the purpose of ROIs, need to be preserved for utility, and hence may need special attention if not already curated into a standard form or code [Miller 2014] [Mayo 2018].

One thing to note about RT objects is that they make extensive use of UID-based cross references to each other. Accordingly, to preserve referential integrity, it is extremely important that they be de-identified together, or in some other manner that assures that when UIDs are replaced, they are replaced with consistent values.

RT data in the DICOM domain is also often linked to other shared clinical data, the risk to re-identification of which needs to be considered in concert with the DICOM structured metadata [Mayo 2016] [Elhalawani 2017] [Kundu 2021] [Kundu 2022].

RT is often used in the treatment of head and neck cancer, so publicly released RT collections need to account for the issues related to reconstructed face recognition discussed earlier. The shape of the entire contour has a significant influence on dose distribution. Without the complete volumetric contour, tools evaluating dose distribution could not be applied with accuracy. This applies not just to intercranial and paranasal sinus regions but also retropharyngeal structures extending up to the skull base. Evaluating dose distribution involves considering tissue density, air volume, photon/proton comparison, etc., therefore full data sets are needed for this.

With a set of contours used for radiotherapy, there may be a representation of the patient's body surface, including the head. Hence there may be some contours or some parts of some contours within RTSS that pose a risk of reconstruction for facial reidentification.

### 1.20.5 Ophthalmology

Ophthalmic images include a broad range of different types, including retinal fundoscopy, slit lamp photography and external photographs, which can largely be handled like any other photographic image, particularly when encoded in DICOM. Some additional modalities like ultrasound and Optical Coherence Tomography (OCT) have different pixel data characteristics but these also have little impact on de-identification. The usual issue of burned-in identification needs to be addressed.

One peculiarity of some ophthalmic images that may be of concern is that the anatomical patterns being recorded have a sufficient component of uniqueness to serve the purpose of biometric identification, such as the retinal vascular pattern [Parrish 2020] [Nakayama 2023] [AAOBoT 2024], or appearance of the iris [Malgheet 2021]. This might become a concern if a sufficiently complete database of such information were to become available to an intruder [AAOBoT 2024].

Ophthalmology and optometry also involve a range of non-imaging modalities and applications. Many of these are supported in DICOM with specific IODs, but more commonly, manufacturers record the output of their devices as laid-out pages of information recorded in PDF format. It has become common to embed these within DICOM Encapsulated PDF form for storage in the PACS. Needless to say, the PDF pages are typically filled with direct and indirect identifiers, which need to be redacted. Recently a structured content mechanism has been added to the DICOM document encapsulation mechanism, so this too requires attention in addition to the normal metadata on traditional DICOM data elements.

### 1.20.6 Dermatology

As for other specialties, dermatologic images may have privacy implications [Scheinfeld 2004] [Kunde 2013] [High 2020]. Many are of course external photographs of the surface features of the body, either closeup or regional, and as such are subject to all the concerns applicable to external photographs previously described, including the exposure of facial features. Burned-in text whether it be a physical label present in the field of view, present on a scanned print or a digital annotation to the pixel data is also a concern. Physical labels (including rulers) used to record identification are not infrequently present in dermatological photographic images [Kazlouskaya 2020] [High 2020] (especially those acquired for research or clinical trial purposes); not being aligned with the rows and columns of the pixel data in pre-specified positions, the information in such labels may be challenging to detect and redact automatically.

A complete set of so-called "whole-body" images may be acquired, either according to a specified protocol as ordinary photographs using a conventional camera, or by an array of cameras or sensors as an automatically reconstructed composite topological 3D surface view [Navarrete-Dechent 2020]. The latter may be challenging to process for de-identification because they may be stored in a proprietary or unusual format, but otherwise pose no specific de-identification challenges.

Dermoscopy (epiluminescence microscopy) [Argenziano 2001] [Fleming 2001] [Caccavale 2020] images are acquired with dedicated devices or camera adapters. The images are close-up views of skin lesions. Unless a digital dermoscope adds burned-in identifying information, the images are generally free of concern. Pattern matching of skin texture remains a hypothetical threat. Dermoscopic images are generally stored in conventional consumer file formats like JPEG, so the general concerns about metadata buried in the header are applicable [High 2020]. Dermoscopic images are frequently accompanied by regional or whole-body images for localization purposes, and these have the usual risks.

More exotic specialized modalities are also used in dermatology [Hamblin 2016], such as OCT [Levine 2017]. The usual concerns about proprietary file formats, identifying metadata buried within consumer format headers as well as burned-in text added to images and PDF files by the manufacturer are equally applicable to such modalities used in dermatology.

Implementation of the DICOM standard for dermatology has been a long time coming [Krupinski 2008] [Madden 2011] [Caffery 2021a] [Caffery 2021b], and though many sites ingest consumer format images and wrap them in DICOM for inclusion in their PACS, few manufacturers of dedicated dermatology devices make an effort to implement the standard or a workflow to automatically ingest demographic and procedure information from the EMR. The incorporation of clinical metadata is important for management and interpretation [Caffery 2020], as well as for feature recognition [Liu 2013]. When dermatologic images are created as or converted to DICOM, the normal procedures already described for the processing of the data elements and pixel data are applicable.

Dermatologists are sometimes reluctant to store the images they or their staff acquire in centralized PACS or EMRs [Milam 2018]. This may be due to privacy concerns given sensitivity over images of the face or of genitalia [Lakdawala 2012] and the lack of appropriately restrictive access controls in many systems. Unmanaged insecure handling of images on physical media and from hand-held mobile devices such as phones also poses a risk [Kunde 2013] [Stevenson 2016].

### 1.20.7 Histopathology

Until the last few decades, digital imaging of microscopy was largely confined to the capture of individual fields through an optical microscope by means of an attached still or video camera, i.e., photomicrography. The procedure was an extension of the process used with conventional photographic film, but was amenable to image processing, which led to the development of a significant area of research. Such images were recorded in a variety of conventional photographic formats, though could also be captured in or converted to DICOM form. Specific DICOM IODs for gross and slide microscopy by such means were added early on [DSC Sup 15]. The de-identification issues related to such material are not significantly dissimilar to those for general photography that they justify a dedicated discussion.

Whole Slide Imaging (WSI), which involves scanning an entire region of a slide at high resolution, was developed later [Pantanowitz 2018]. An appropriate IOD was added to the DICOM standard [DSC Sup 145]. WSI using visible light and fluorescence for traditional brightfield microscopy and immunohistochemistry now sees significant application for research and is increasingly being deployed for clinical use as the technology advances. The secondary re-use of clinically acquired images raises the same de-identification concerns as any other source of images, but the nature of the set of images acquired for each scan, the characteristics of the physical material imaged, and the manner in which they are encoded are quite distinct. The ethical issues of de-identification when WSI are used for training of computational pathology applications are recognized [Sorell 2021]. The need for a comprehensive approach to handling de-identification for WSI on a large scale has been described [Schüffler 2021a] and tools to handle standard and proprietary formats are becoming available [Bisson 2023].

### *1.20.7.1 Whole Slide Image De-identification*

The process of acquiring a WSI from a glass slide produces a set of images, typically consisting of:

- an image of the slide label
- an overview or macro image of the entire slide, including all or part of the label and the region of the slide containing tissue
- one or more images of the region of the slide containing tissue

The label and overview images are typically low-resolution and each encoded as a single image frame. The tissue regions are acquired at high-resolution, very often split into tiles [Silage 1985] or strips due to their very large size, and frequently accompanied by down-sampled derived images comprising a multi-resolution pyramid of tiled layers to facilitate rapid zooming and panning by a viewer [Harris 2001] [Aperio 2008], although the so-called "virtual microscopy" experience can be accomplished by other means of encoding, such as by using wavelet decompositions [Ferreira 1997] [Hulsken 2016]. The following is something of an oversimplification based on a typical size glass slide and may not be sufficient to describe less common patterns, such as applied to larger slides for whole mounts, or where tissue from multiple specimens is included on the same slide (a tissue microarray (TMA).

As with any other type of image, identifying information can leak through four mechanisms:

- file or folder name information (and other filesystem metadata properties)
- image metadata embedded within the image file
- burned-in information within the image pixel data (whether added or in the original subject)
- metadata encoded separately but included in the collection (e.g., in a spreadsheet with the images)

Those issues that are specific to WSI or types of WSI will be discussed. It is particularly important to consider images other than that of the slide label for potential identity leakage, whether multiple images are embedded within a single file, or in separate files [Zarella 2022]. The question of matching similar WSI [Holub 2023] is not considered here, since the risk models are the same for all image types as previously described, though the matching techniques may differ.

#### 1.20.7.1.1 Slide Label Image

De-identification concerns arise primarily from the slide label. In the laboratory, the slide label is used to identify and manage the workflow associated with the handling of the slide [Brown 2015], whether the purpose of acquisition is for research or for clinical use. The slide label may contain only a single identifier of the slide in the form of an alphanumeric code, a barcode, or both, or it may contain additional information identifying or describing the slide, specimen, accession, procedure (including stains and dates) or subject. When slides are sent to another organization, the label of the sending organization may be replaced (covered over) with the label of the receiving organization. When a clinical slide is reused for research purposes (e.g., scanned at a contract research organization (CRO)), again a different label may be applied. Since a physical asset is being scanned, the slide label might be physically redacted (e.g., with a marker pen or opaque label) before scanning, though the effectiveness of such redaction may not be complete.

The slide label may not always be in the normal location at one end of the slide. Sometimes there may be labels at both ends of the slide. Even if it is present in the normal position, a label may be larger than usual (and expected by the software). Sometimes there may be additional non-standard size or shaped labels present in the region that is normally occupied by tissue. In such cases, only the normally positioned label will be scanned and extracted into the digital slide label image, but the other labels will appear in the overview or macro image and may conceivably be encountered in the tissue image.

Even if part of a slide label appears dark to a human observer at default settings, e.g., appears to have been redacted, image enhancement may still reveal text that was not initially apparent.

The slide label contents are equally critical to the handling of digital images of the slide. Accordingly, the scanner not only photographs, separates, and records the slide label, but may also perform OCR and barcode recognition on its contents [Englert 2024], and encode the extracted information in metadata, which may be embedded in the stored image, used to name the file and folder in which the image is stored, or be used in some other manner to persist the association of the digital asset with its identification.

The bottom line is that for de-identification purposes, as long as the digital images are adequately tracked by accompanying metadata, which itself has been or will be pseudonymized or de-identified as appropriate, the label image can generally be completely discarded. The only reasons to retain it, if it is known to be safe in the sense that it contains already redacted or pseudonymized information anyway, are for quality-control purposes (check that the metadata matches the slide), and to perform research into scanned label image processing (e.g., barcode or text OCR [Englert 2024]). Tools exist to selectively remove the label images embedded in multi-image formats [Gilbert label], and DICOM WSI label images are clearly identified as such. If labels are retained, any barcode present needs to be decoded and checked to assure that it does not contain identifiable information. When the workflow would benefit from the presence of a visual barcode with pseudonymized identifiers and de-identified descriptors, new label images can be synthesized and inserted.

#### 1.20.7.1.2 Slide Overview Image

The overview or macro may contain part or all of the traditionally positioned slide label, as well as any additional unusually placed labels as described. It may also contain any physical annotations made on the slide, such as with an ink pen to outline tissue or regions of interest, and theoretically this can include handwritten comments [Schüffler 2021b]. Though it would be unusual for these to contain identifying information, it is possible. Techniques have been developed to detect and remove ink from slide images, albeit not specifically for the purpose of de-identification [Ali 2019] [Maleki 2020] [Jiang 2020]. It may be desirable to preserve some annotations that have clinical significance for interpreting the pathology. Additional text is sometimes present that is clearly non-identifying, such as the etched-in manufacturer of the glass slide.

Conversely, the overview images are extremely important, especially in a clinical context, to allow the user to assure that all pieces of tissue physically present on the slide have been scanned [Atallah 2021], as well as to provide a visual context for the tissue region image(s) and navigation thereof.

So, being useful, but posing some risk, if a means can be implemented of checking the overview images for the presence of labels or burned-in text, and redacting it, whether manually or automatically, the images can be retained. Otherwise, they need to be removed.

#### 1.20.7.1.3 Tissue Images

The images of the tissue regions that are captured at high resolution, if they contained only tissue or empty space, would not pose a risk. However, since they are generally rectangular in shape and may capture significantly more than just tissue, and since they are defined before scanning either automatically or by a human operator, there is some risk of including annotations or unusually placed labels as described earlier.

The high-resolution images are generally already accompanied by down-sampled layers encoded as a pyramid, even to the extent of including a single tile thumbnail image that is representative of the entire area of the slide covered by the pyramid. It may be possible to review only the top layer (or higher layers) of the pyramid for the presence of labels or burned-in text, without having to analyze every tile of the high-resolution layer, which would be time consuming or computationally expensive.

Though tools are not yet generally available for application to WSI, the same burned-in text detection and recognition techniques that may be applied to other types of images are obviously applicable. Ink removal techniques have been described in the slide overview section. Having detected text for redaction in a lower resolution image of a pyramid, it is necessary to have tools that propagate the redaction region to the other resolution layers of the pyramid. Human-readable text detectable in low resolution images, when propagated into the highest resolution layers, whose individual tiles are of small size and whose individual pixels are microscopic (literally), text may no longer be detectable within a tile considered alone. Another consideration is that whatever redaction process is applied, it should preferably not involve decompressing and recompressing every tile of all layers, only those that need redaction, since otherwise further loss is caused, as discussed for ultrasound.

#### 1.20.7.1.4 Slide Metadata

Unlike for many medical images, there is far less standardization of file formats and metadata in the WSI community [Kayser 2008], particularly for research purposes [Swedlow 2021]. Interoperability has traditionally been dealt with by using libraries that can read multiple formats rather than standardizing on a single format [Goode 2013] [Moore 2015]. Even to the extent that alternatives to DICOM have been adopted for WSI, such as OME-TIFF [Goldberg 2005], the scanner vendors persist in using their own proprietary formats. Fortunately, many of these formats are TIFF-based and follow similar patterns. For de-identification, significant effort must be put into developing format-specific tools to detect and remediate format-specific metadata. An alternative is to convert the images into a common format (DICOM, OME-TIFF or generic TIFF) prior to de-identification [Bisson 2023].

Proprietary WSI files do not typically contain a lot of patient identifying or descriptive information within their metadata. Managing such files may depend almost entirely on their file name and location in a file hierarchy, rather than anything embedded within them [Zarella 2022]. Though extremely disadvantageous from a reliability and scalability perspective, this does mean that the risk of identity leakage from direct patient-related identifiers is relatively low. However, to the extent that dates and times related to the patient are present [Taylor 2022], for example, what little embedded metadata is present cannot be ignored. To state the obvious, the corollary of this is that file and folder names need to be changed to reflect the pseudonymous replacement identifiers.

The general principles that have been described for rule-based de-identification earlier in the report are equally applicable to WSI. So too are the specific concerns related to the detection and removal or identifiers included in unstructured or semi-structured descriptive text.

Proprietary and generic TIFF-based WSI are generally well-structured. The TIFF standard tags [Aldus] describe in detail what information is contained therein. Non-standard proprietary tags may need some reverse engineering if not documented by the vendor; otherwise, they need to be removed. There is not currently a readily available database of all vendors' proprietary metadata fields with an indication of which are safe from identify leakage or not. Some insight can be gained from inspecting the open-source tools that can read such formats.

Some extensions intended for microscopy do not use the TIFF tag mechanism, but instead encode other structured format content into single TIFF tag values. In particular, the TIFF Image Description tag value often contains unstructured, semi-structured, or even XML text metadata. It may be simple or extremely detailed, for example, in the case of OME-XML [Goldberg 2005]. Though this information is usually mostly related to the encoding of the data or the manner of acquisition, it may contain identifiers of patients, other persons, dates and times, device, and institution information. This text needs to be cleaned or removed as appropriate. It is frequently extracted into structured metadata by reading libraries and applications. When proprietary format images have already been converted, proprietary information may have been extracted and stored in other fields. E.g., for SVS images converted into OME-TIFF, the metadata extracted from Image Description may be re-encoded as original annotation XML tags in an OME-XML description. Accordingly, a very thorough review of the multitude of places the same information may occur is required (e.g., for dates [Taylor 2022]). Further, if the images are to be left in their original format but selective tags removed or replaced, though there are generic libraries and tools for reading TIFF-based formats, for example, there are not many tools that fully automate the removal of specific offending information. An example of the need would be editing (rather than completely removing) an SVS Image Description tag value to selectively remove identifying metadata but retain important acquisition information such as the spatial position and acquisition parameters, in such a manner that SVS software would still recognize the information correctly. The editing task may be complicated by the need to update binary offset pointer tables to preserve the structural integrity of the file.

If images are converted into DICOM prior to de-identification, then all the usual principles of DICOM de-identification can be applied unchanged. This will automatically address decision-making and implementation of metadata de-identification in the usual modality-agnostic manner. Any text or private data elements that contain information propagated from the proprietary format will need to be cleaned or discarded. Label, overview, and tissue region images will be clearly identified, and can then be discarded or selected for cleaning as appropriate. If dual-personality DICOM-TIFF files [Clunie 2019] are used, care must be taken to assure that the metadata in both the DICOM and TIFF parts of the image are addressed; generally one will be derived from the other automatically during writing, so this may not be a practical concern.

The image processing principles applied to detecting the presence of labels or burned-in text, and redacting them, are equally applicable regardless of the format in which the WSI is encoded. The issue is the availability of suitable libraries to read, edit and write the necessary input and output formats.

Since WSI are often encoded in compressed form, e.g., as lossy JPEG tiles, care should be taken to ensure that identifying metadata is not buried within the compressed pixel data bitstream, as previously discussed for such formats. In practice, this has not been encountered, since the scanner vendors try to keep the compressed bitstreams as small as possible, but is a theoretical concern that would be easily mitigated by scanning all tiles for unnecessary marker segments.

## 1.21 Artificial Intelligence and De-identification

Artificial Intelligence (AI) impacts de-identification both as a consumer of de-identified images as well as a potential tool for improving the process of de-identification.

The application of AI to text recognition for de-identification purposes has previously been discussed in the context of unstructured text as well as burned-in text.

There has been research into the generation of synthetic data as a replacement for real data [Shin 2018] [Morrison 2022], but that is not germane to the actual performance of de-identification.

## 1.22 Reports, Documents and Annotations

### 1.22.1 Overview

Medical images are often accompanied by and released with other information. Tabulated metadata has already been alluded to as requiring the same attention as metadata contained within the image files. Another category of information includes qualitative or quantitative statements about the images themselves, the manner in which they were acquired, or their relevance to the patient's condition. The use cases are essentially unbounded but may be constrained by the purpose for which the data is being shared. I.e., the simplest method of assuring that such information does not create a reidentification risk is not to include it in the released data in the first place. However, annotated images are generally considered to be much more useful than the images alone [Torres 2012]. Accordingly, this report will consider the implications of various categories of accompanying data.

Note that analysis results generated exclusively from already de-identified information are generally not a concern, since they do not contribute to any pre-existing re-identification risk.

### 1.22.2 Clinical Reports (Narratives)

Traditional reports about images, whether they be radiological or pathological, or in the form of clinical notes about images obtained in the course of some clinical activity, are relatively unstructured and are often dictated as prose (narratives), albeit sometimes with the use of templates to constrain the organization of the text.

There has been a large effort devoted to completely general approaches to processing unstructured text, whether they be for health care specific documents, or of general applicability without regard to industry. The field of extraction of semantics from unstructured content (+/- de-identification thereof) has been considered in a healthcare setting [Sweeney 1996] [Ruch 2000] [Douglass 2004] [Dorr 2006] [Uzuner 2007] [Neamatullah 2008] [Aberdeen 2010] [Taira 2010] [Stubbs 2015] [He 2015] [Dernoncourt 2017] [Langarizadeh 2018] [Norgeot 2020], for radiology reports [Haug 1990] [Friedman 1994] [Demner-Fushman 2016] [Steinkamp 2020] [Tenenholtz 2020] [Mozayan 2021] [Casey 2021] [Horng 2022] [López-Úbeda 2022], radiotherapy notes [Zhou 2021b] and pathology reports [Berman 2003] [Gupta 2004] [Beckwith 2006] [Chen 2019] as well as for generic data loss prevention through named entity recognition [Taira 2010] and other forms of natural language processing and understanding. Not only are there extensive descriptions in the literature but also support in commercial, research, and open-source tools.

As discussed earlier, these general approaches to unstructured text, whether based on hand-crafted patterns or machine learning techniques, can also be used on text fields within image metadata as well as OCR'd burned in pixel data text [Kopchick 2022].

The manner in which the report is encoded in the supplied data to be de-identified and in the de-identified result varies. In typical clinical practice reports are exchanged as the plain text payload of an HL7 V2 message [IHE RD], and extracted text may be supplied. This is not an ideal format for sharing more broadly, so curation may involve recoding into some more relevant format, such as a DICOM SR. Regardless of the supplied or resultant format, the same principles apply to the de-identification of the text payload. Report formats will be further addressed under the more general categories of Annotations and Documents.

### 1.22.3 Annotations

In this usage, we consider annotations to include categories, labels, notes, measurements, or similar relatively brief, focused aliquots of information, with a specific focus, such as a pixel, a region of interest (ROI), an image, a series of images, or an entire imaging study, patient, or specimen. Typical examples would be measurement of the longest diameter of a nodule, segmentation of an image into tissue types, selection of a particular image as being of key interest for some specified reason, and categorization of a series or study as being of a particular type. In DICOM, such annotations are encoded in various different non-image objects, which share the same information model (patient, study, series and instance) and have that composite context encoded in every instance but differ in their payload.

DICOM Structured Reports (SRs) encode name-value pairs (content items) of annotations where the name is always coded but the value may be of various types. Like DICOM data elements, these name-value pairs encode structured metadata, and can be treated in a similar manner, i.e., subjected to rule-based decisions with respect to the risk that their content poses and de-identified accordingly. DICOM PS3.15 Annex E contains tables of coded concepts with appropriate de-identification actions specified [NEMA PS3.15 E.1]. Just as DICOM instances are constrained by their IOD, so too are DICOM SR content items constrained by templates defined in [NEMA PS3.16], so care needs to be taken not to invalidate the compliance with the relevant template during de-identification of the SR content tree. As with data elements, the content items are strongly typed, and the same principles apply during de-identification.

SRs may describe locations or regions by explicit 2D or 3D contours or reference other objects that provide location such as RT Structure Sets (RTSSs) and pixel-based (raster) or surface-based segmentations. Segmentation objects may stand alone without SRs and contain their own coded and descriptive information. It is important that the de-identification process be performed so as to maintain referential integrity (especially with respect to UID replacement), so that the connection between an annotation and what it is annotating is not lost (e.g., a reference to a frame or image in the case of a 2D image-relative contour annotation, or to a Frame of Reference in the case of a 3D contour annotation, or to the instance of a segmentation object that defines an ROI)).

SRs may also include large amounts of text and indeed can include an entire clinical report or note, which needs to be treated as any other body of text from a de-identification perspective, and either be cleaned or removed or replaced as appropriate.

The annotation objects described so far are intended to be machine understandable, but there are other mechanisms that were designed to present visual information to humans. These include DICOM Presentation States as well as various forms of standalone or embedded bitmapped overlays. Such encoding of annotations can be more challenging to categorize from a de-identification perspective since they are inherently unstructured. Text identified as such can be processed like any other unstructured text, but bitmaps and vector graphics are more challenging to classify and may need to be routinely removed in the absence of some OCR-like mechanism to interpret them. E.g., a hand-drawn outline, squiggle, or arrow may be difficult to distinguish from cursive text that contained identifying information. The same theoretically applies to contours in more structured objects like SRs and RTSSs, but to the extent that they are accompanied by categorizing codes and a specific purpose, they may be less risky. One approach is to have these reviewed by a human to determine their disposition.

### 1.22.4 Documents

In this category we will distinguish what an object contains as opposed to how it is encoded and consider both from a de-identification perspective. The clinical reports and annotations described so far can sometimes be considered as documents in a general sense, and indeed they may be explicitly encoded as Encapsulated Documents in a DICOM sense. For example, a clinical report may be encoded in Portable Document Format (PDF) and either sent that way or wrapped in a DICOM header and encoded as an Encapsulated PDF. The PDF format contains its own explicit metadata encoding distinguished as such from the payload, which is relatively complicated to decipher. Should such documents be included in a collection for release, diligence in redacting not only the payload [Kuzmak 2019], but any embedded or DICOM metadata is required. Manual redaction tools are available, and some commercial data loss prevention tools intended for other industries may be useful [CaseGuard]. Care should be taken to assure that any identifying text in the PDF is actually removed (redacted) rather than just covered up, since otherwise the underlying text is recoverable and sometimes visible.

DICOM also allows for encapsulation of a relatively short list of other file types, which are also referred to as being Encapsulated Documents even though they are not really documents in a general sense. HL7 XML-based CDA [Dolin 2006] documents of various types, including diagnostic imaging reports [NEMA PS3.20], may be encountered, though uncommonly, and these too are supported in DICOM in an Encapsulated CDA form [NEMA PS3.3 Encapsulated CDA].

Three-dimensional printed model formats including STL and OBJ are also supported as DICOM Encapsulated Documents, and each of these needs to be addressed specifically by an expert in the appropriate format with respect to de-identifying the metadata and the payload. E.g., an identifying serial number might be embedded in the 3D data, for example, so that it appears on the printed physical model. Beyond warning that these objects need to be detected and sequestered (e.g., based on their DICOM SOP Class UID), further discussion of these types of encapsulated objects is beyond the scope of this report.

Considering the more general concept of a document, without regard to its encoding, a specific type of concern is a scanned document that was the paper request form for an imaging study. In practices without paperless ordering, these scanned documents were often included in the study with the other images. Other paper artifacts such as signed consent forms are sometimes encountered. In general, this category of object is not usually useful to include in collections released for secondary re-use, though they can be redacted like any other document, through OCR if necessary, or by manual curation. Such scanned documents may be encountered most commonly as DICOM Secondary Capture [NEMA PS3.3 Secondary Capture] or Encapsulated PDF [NEMA PS3.3 Encapsulated PDF] objects, and robust de-identification procedures need to have a means of detecting them either to exclude them or to handle them specially and subject them to an appropriately high level of quality control.

## 1.23 Evaluation

Distinct from matters of re-identification risk are questions of de-identification system performance; i.e., how well a de-identification system performs with respect to pre-specified requirements to remove or replace specific data elements or text values.

For structured data this should be reasonably straightforward when the original values are available for comparison. Given a list of single structured fields that are purported to be direct or indirect identifiers, one can easily ascertain that they have all been processed. For the requirements specified in the DICOM standard to remove or replace certain data element values [NEMA PS3.15 E.1], input containing objectionable values can be fed to the system and compared with the output to determine that they have indeed been removed or replaced. The generation of such a check can be automated using the tables in the standard. E.g., one can mechanically check that the Patient's Name field is indeed empty or has been replaced with a dummy value (that is recognizable as such, as opposed to a genuine, but incorrect, name). Conversely, data elements that should be retained with their original values can be similarly checked. When the expectation is that a value should be replaced, it may or may not be safe to assume that if any change has occurred, then it is sufficient.

Checking that the field has been changed to an appropriate rather than inappropriate value, or indeed that it has been sufficiently changed, may be a non-trivial task. E.g., it is no longer considered acceptable practice to replace a Patient's Name with their initials, though this was once common practice. If values are being "cleaned", i.e., processed to remove unsafe information and retain safe information, then the check is more challenging. It is insufficient to simply check for any change. Rather the changes must be evaluated as satisfactory. One approach is to define the expected output value specifically and expect an exact match. Any deviation is flagged for further inspection and may or may not be found to be satisfactory. Some mechanical tests beyond change detection can also be performed. For example, just as an ensemble of different text de-identification tools may be applied in succession, which together may perform more thoroughly than individually, one pattern recognition tool may be used to search for suspicious residual identifiers in a purportedly de-identified output that was processed by a different pattern recognition process.

Value comparison is further complicated by application of de-identification rules or methods that do not produce deterministic output. For example, UIDs may be replaced with different values on different runs, depending on the technique used.

The result of such checks may result in extensive output with considerable repetition, particularly when many test images are being evaluated. Manual review of the output may be simplified by sorting the output for unique patterns.

DICOM conformance of the output should be checked. The result should be no worse than the conformance of the input. Improvement is not expected, but the de-identification process should not generate invalid values. Particular care should be taken to check that inserted fixed-length replacement values do not exceed the maximum permissible length or result in violation of the data type specified for the VR of the data element. E.g., a common error is to replace dates with strings that are not valid dates (like "DEIDENTIFIED") and this should be detected and recognized as a failure. While such strings may serve to communicate the reason (trigger) for the replacement, that can be accomplished by creating separate logs rather than invalid values in DICOM data elements.

Individual errors can be enumerated, counted, expressed relative to the number of images de-identified, and tracked or monitored as improvements or changes are made. It is difficult to come up with a single numeric score that represents the overall "quality" of the de-identification in a meaningful way. Such a metric would be important, for example, to use as a comparison between competing methods in a challenge scenario [Uzuner 2007] [Maier-Hein 2018], or judging competing bids. If for no other reason, some errors are more important than others, and any weighting of different categories of errors to produce a single metric is somewhat arbitrary. Arguably, a metric and its weights may be unfair to one algorithm when being compared against another. Further, the numeric output may be difficult to normalize in a manner that accounts for the volume and the diversity of the input (or lack thereof), in that there may be an excess of some types of triggers and a paucity of others. This is especially true when trying to include performance of rule-based de-identification of structured data elements, extraction and redaction of unstructured text, and detection and remediation of burned-in text in a single metric. It might be appropriate to define a small number of separate metrics and draw conclusions from them without trying to combine them.

A canonical test set of input data designed to exercise a range of features may be necessary to allow meaningful comparisons. As with challenges in other fields, it may or may not be necessary to release a shared training set separate from a sequestered testing set used for evaluation, and it may or may not be desirable to release all the data after the challenge is complete to provide transparency into the evaluation process. The manner in which the results of a challenge or competition are reported has been demonstrated to be important and standard reporting practices are being defined [Maier-Hein 2020]. These are likely to be equally important for de-identification challenges or competitions. Given the difficulty of gaining access to images with actual patient identifying information still present, some sample images with synthetic identifying information have already been made available for testing [Rutherford 2020]. More images from a broader range of modalities and applications illustrating a greater variety of identification test scenarios will be necessary to serve as a sufficient corpus in the future. In addition, as automated de-identification tools become more sophisticated, and for example, able to distinguish real from synthesized information, and remove the former but pass the latter, more sophisticated tests may be required. For example, issues with using synthetic street addresses have already been observed with one platform [Kopchick 2022]. For the purpose of evaluating rule-based de-identification, the test data sets can be enriched with problematic information, as long as the evaluation metric is not affected by the balance of features. Special attention to private data elements is required in synthetic data, to challenge overly simplistic configuration mechanisms (e.g., failure to use the creator ID, assignment in uncommon data blocks, etc.). Issues of bias in the test data set such as due to limited demographic representation also need to be addressed [Xiao 2023].

A completely different approach that could also be used as a metric in a challenge is to use quantification of residual re-identification risk, or reduction of such risk after de-identification. An appropriate SDC model (or set of models) for a selected threat could be used to evaluate the extracted structured metadata and the de-identified direct and indirect identifiers therein, potentially using existing tools. This would not address the detection of burned-in text or image features. Data sets used would need to be statistically balanced to represent a valid sample of the population (in terms of unique records and equivalence class sizes), rather than being enriched, to avoid validating the underlying model producing the risk metric.

An interesting challenge in the privacy space, albeit for voice rather than images, which attempts to define metrics to quantify both privacy and utility is the VoicePrivacy challenge [Tomashenko 2020] [Tomashenko 2022].

## 1.24 Motivated Intruder Attack

A special case of evaluation, which is specifically related to re-identification risk as distinct from conformance to requirements, is the motivated intruder attack [ICO 2012] [Garfinkel 2016]. In essence, this involves arranging for an intruder who is *"reasonably competent"* to attempt to reidentify de-identified records within the collection, before release. This approach is particularly useful for public releases and may identify unexpected ("unknown unknown") threats. Even if the motivated intruder cannot verify any matches obtained in the attack, the confidence in the match may be a useful surrogate [Elliot 2016b].

Such tests may need to be repeated as new data is de-identified, as available comparison data sets and methods of attack evolve, and as new evidence regarding the capabilities of computational tools that can facilitate re-identification emerges. Health data is generally considered to be at high risk and intruders potentially well-motivated [Trustwave 2017]. This may theoretically extend to include re-identified medical imaging data and/or the additional data accompanying the images, though as far as we know this has not been documented or estimated.

## 1.25 Quality Control

Regardless of whether de-identification is being performed on a sporadic basis in an ad hoc manner by individual persons, or as an industrial process on a large scale in an automated manner, there is a need to confirm that the process has been successful. Success in this respect means that the pre-specified requirements have been satisfied, and implicitly or explicitly, that a pre-specified re-identification risk threshold has not been exceeded.

Herein we will focus on industrial-scale, high-volume de-identification, and the corresponding quality practices and procedures applicable to any similar manufacturing-like activity. This is not to say that individual persons performing de-identification on their own cannot adopt a "personal de-identification process", scaled down perhaps, but designed to achieve similar goals (with apologies to Humphry [Humphrey 1997]). Even something as simple as a checklist [Degani 1989] to follow after de-identification may be better than nothing and may serve to provide an explicit confirmation function [Lingard 2005].

The de-identification of medical data on an operational scale is an industrial process. As such, well-established industrial practices for the documentation, deployment, execution, maintenance, monitoring, improvement, and management of processes should be applied. It is beyond the scope of this report to select or recommend any quality system or method, particularly as these evolve over time. However, some general principles and aspects that deserve special attention will be considered.

The initial deployment of a de-identification process requires special attention, particularly for one with configurable or customizable features. All tools and processes to be used need to be qualified in some manner as fit for purpose, preferably against clear, unambiguous, pre-defined and objective success criteria. Ideally such testing would be deterministic and complete, in the sense that every test run gives the same result and every pathway possible is tested. Neither may be possible in practice, but at the very least, some quantitative estimate of both completeness testing and of residual re-identification risk should be made. This should be documented as adequate for the context of use when compared to the organization's tolerable risk threshold, the establishment of which should be justified. Critical to any such assessment is identification of a credible threat model.

We distinguish those activities that are performed 100% of the time on 100% of the data and which are therefore part of the de-identification process itself, from those that are performed selectively or on a sample and represent a quality check. This is true regardless of whether the activity is performed by a human or a machine. For example, some checks, such as the successful detection and removal of burned-in text might require a human observer, or perhaps the application of an extremely computationally expensive machine algorithm that cannot practically be performed on all the data. The state of the art in burned-in text removal currently still requires a "human in the loop" due to an excessively high failure rate [Kopchick 2022], but that may not remain the case as algorithms improve.

To reduce cost, quality checks should ideally be performed selectively or on a sample. Worst case though, and arguably still the safest and most conservative state-of-the-art approach, is a 100% check of all at-risk fields by, at least one, human expert. Even then, humans are imperfect at manual repetitive tasks such as this, as well as being prone to fatigue. Though 100% human checking is expensive and time-consuming, it does serve to mitigate (though not eliminate) the potential of even a single re-identification occurrence, which might undermine the credibility of the service. The thoroughness of quality checks to apply is a cost-benefit decision, which ideally would be made considering a robust quantitative assessment of the risk, as well as measurement of the performance of any automated tools that can be brought to bear. Even if checks are applied to a smaller fraction of the data than 100%, at least some checking is always required, at least some human checking is strongly recommended, and normal statistical sampling methods, perhaps accounting for higher risk subsets, are appropriate.

Every single action performed by a de-identification tool is potentially amenable to a quality check. However, some entail more risk than others, either in terms of the hazard (consequence of failure) or likelihood of occurrence. An element of judgement is required in terms of selecting which checks to pursue. The decision may be guided by experience with which actions have succeeded or failed in the past, as well as the complexity of the individual task. As with any quality-related activity, the gathering of empirical data to guide such decisions is imperative. We do not intend to enumerate (again) every change that a de-identification process may implement. Obviously, careful attention should be paid during quality checks to assure that all direct identifiers have been removed or replaced in structured and non-structured data, as have indirect identifiers established as being of concern, in addition to any descriptive and burned-in text.

The burdensome task of human quality checks of plain text fields can be alleviated by pooling techniques. For example, one can separate strings of short length (by sentences, phrases, or keywords, or between various delimiters) and then sort them uniquely. That way, far fewer strings need to be examined (since medical imaging metadata values are highly repetitive). Suspicious strings that may be identity leaks stand out visibly more readily when sorted. The shorter sorted lists may be scrolled through by a human more quickly.

A recognized challenge to the practicality of human review of all image pixel data is their sheer volume, coupled with the need to adjust contrast settings to assure that all embedded information is visualized. Two approaches to mitigating the work effort may be considered. First is stratification of the collection to target those images at greater risk; e.g., those from known dangerous modalities like ultrasound, or which have structured metadata data element values indicating that they are secondary capture or screen shots or contain overlays or vector graphic annotations. This may employ a similar process to the selection of images for redaction in the first place [Clark 2013]. Employment of this tactic should be subject to a documented risk analysis, preferably based on empirical data. Another tactic is to use a tool that allows multiple images to be evaluated simultaneously, such as by using cine mode [Clark 2013], or to create a maximum intensity projection (MIP) for a set of parallel cross-sectional slices [Bennett 2018b] [Jarosz 2018]. Regardless of the techniques employed, the state of the art for this task remains a 100% check by at least one human, who reviews every image rendered on the screen [Prior 2017].

Changes to the de-identification process may occur, and the nature of the input data may evolve over time, so the de-identification process needs to be monitored over time in a manner that is able to detect changes in its performance. It should go without saying that a rigorous change control process should be implemented. That way any changes in processing or data can be anticipated and validated before deployment, making the impact of such changes predictable. Monitoring of the performance then serves to confirm successful changes as well as identify unanticipated consequences, including regression of behavior. During routine maintenance, the consequences of a configuration error with an undesirable effect are potentially severe, so extreme caution should be taken to also test and check the effect of modifications before deployment. A checklist-based approach, and successive checks by different individuals may be useful. Automated regression testing is recommended, using the same suite of tests as used during initial deployment, augmented with additional tests of new or changed features.

Currently the literature is relatively silent on the efficacy of de-identification processes and the actual need for quality checks in practice, though articles addressing tools for quality control of curation and de-identification are starting to appear [Kosvyra 2022]. Ideally, experienced practitioners would report on how often what types of issues are found, so that the rest of the community could base their risk management and process design decisions on objective evidence.

We will not specifically address quality management aspects of de-identification, including quality improvement, but as with any other large-scale operation, no process is perfect, there is always an opportunity to improve, and the usual approaches are applicable [Kruskal 2009].

## 1.26 Tools

The scope of the task group's work does not include performing a comprehensive review of all available tools. Nor will any specific recommendations of individual tools be made.

There are many tools described that perform DICOM image de-identification, some of which are freely available, open-source or commercially available, some of which have a user interface and others are command line [Toms 2006] [Bland 2007] [González 2010] [González PrivacyGuard] [DICOM Confidential] [Lien 2011] [Li 2011] [Suever 2011] [AnonMed] [Archie 2012] [Newhauser 2014] [Doel 2017] [DICOM Anonymizer] [Veldhuis] [Haumont] [CHOP-DBHI] [Jodogne Orthanc Anon] [Clunie DicomCleaner] [Cooper] [O'Dell] [Bazargani] [IBM UDIP] [Rubo] [Neologica] [Santesoft] [DicomSystems] [SeeMode] [Rosenfield] [Google Deid] [Google Redactor] [Wiggins 2019] [Pure Image] [BMD 2022] [Microsoft DICOM] [Sengupta 2019] [Shahid 2022] [Kitware]. The references provided here are by no means an exhaustive list. Some tools and services may no longer be available, supported or maintained, or may be development efforts that have been abandoned. Most use a rules-based denylist/blacklist approach. Some do not do a particularly thorough job of de-identification [Patel 2014]. Some are configurable with profiles and options and others at the individual data element level through scripts or by means of a user interface, which may allow for user review of sufficiency of de-identification [Shahid 2022].

Most of these tools are intended for interactive use on a small scale, though some are designed to run in bulk. Some provide background servers to process incoming images in real time. The tools support a variety of operating systems and are provided in various programming or scripting languages. A few tools are modality specific (e.g., for ultrasound [SeeMode] [Rosenfield]). Some provide means of manually or automatically redacting burned-in text [Sochat 2017] [Google Deid] [Wiggins 2019] [Pure Image]. Some general purpose DICOM toolkits also provide de-identification utilities [Malaterre gdcmanon] [DVTk Anon].

A number of large-scale projects [Aryanto 2011] [Aryanto 2012] [Clark 2013] [Granite 2013] [Kalpathy-Cramer 2014] [Aryanto 2015] [Bennett 2018a] [Erdal 2018] [Mesterhazy 2020] make use of the Clinical Trial Processor (CTP) [RSNA MIRC CTP] as a script-driven tool for bulk processing. Others [Yi 2021] use different tools like Orthanc [Jodogne Orthanc Anon], and some build their own tools for the purpose [Patel 2014] [Marques Godinho 2017]. Some general purpose scripted DICOM editing tools are also intended to be configured for de-identification either server [XNAT DicomEdit] or client side [Cho DicomEdit].

Many image viewers and PACS have built-in basic de-identification features [Shamshuddin 2014]. These may be very limited and give the impression of performing de-identification but not be particularly thorough [RCR 2019], without extensive customization [Roddie 2016], if even possible. We did not have access to any recent evidence assessing built-in de-identification features of modern commercial image viewers and PACS. A thorough study of such features would be appropriate.

Users are cautioned to be wary of descriptions and comparisons of image de-identification tools that may be found in the scientific literature [González 2010] [Lien 2011] [Archie 2012] [Fetzer 2014] [Aryanto 2015] [Chane 2020] [Shahid 2022]. These may provide a useful inventory and suggest evaluation strategies. They may also oversimplify the problem or be less comprehensive in their evaluation than might be desirable. As a result, they can be misleading [Clunie 2016b] [Aryanto 2016] [Kundu 2020b]. Instead, we recommend that users perform their own evaluation of a particular tool's claims by reviewing its documentation to establish feasibility, then thoroughly testing its performance in a range of configurations against objective criteria, such as the DICOM profile and various relevant named options. Sample images with synthetic identifying information are available for testing [Rutherford 2020]. User's may wish to compare the results of different tools on the same data sets to gain insight into their performance. De-identification is not a simple problem, and use of a simple tool may give undesirable results.

We remind readers that no known image de-identification tools address the matter of statistical disclosure control and re-identification risk. This is a deficiency that may eventually be rectified. In the interim, if patient characteristics are retained, being potential indirect identifiers, it is necessary to augment existing tools with additional processing steps.

## 1.27 Conclusions

In this report we have summarized the best practices for medical image de-identification based upon the experiences of the task group members and our review of the pertinent literature, and documented the relevant background material. Where knowledge gaps exist, we have made specific recommendations for future investigation. Where alternative approaches are feasible, or appropriate for different sectors, we have tried to fairly compare them. We recognize that our conclusions may be somewhat cancer-research-centric. Where there is a lack of consensus, we have tried to provide a balanced viewpoint, though we recognize that there may be areas where disagreement persists.

We recognize that our conclusions may not be completely satisfying for those seeking a turn-key off-the-shelf solution for what is a relatively complex problem. We also emphasize the importance of continued vigilance as the types and sophistication of threats evolves, and the potential need for a quantitative approach to risk analysis. It is not our intent to convey a nihilistic message, that public image data sharing is impractical. but rather to emphasize that the immense value of such sharing justifies the investment of the non-trivial effort required to balance protection of privacy with retention of scientific value.

## 1.28 Abbreviations and Acronyms

ACR - American College of Radiology

AI - Artificial Intelligence

BIDS - Brain Imaging Data Structure

BLOB - Binary Large Object

CBIIT - Center for Biomedical Informatics and Information Technology

CAD - Computer Aided Detection

CDA - Clinical Document Architecture

CDISC - Clinical Data Interchange Standards Consortium

CRO - Contract Research Organization

CSV - Comma Separated Values

CT - Computed Tomography

CTP - Clinical Trial Processor

DCE - Dynamic Contrast Enhanced

DICOM - Digital Imaging and Communications in Medicine

DNG - Digital Negative

DSC - DICOM Standard Committee

DVH - Dose Volume Histogram

ECG - Electrocardiography

EEG - Electroencephalography

EHR - Electronic Health Record

EMR - Electronic Medical Record

EXIF - Exchangeable Image File Format

FDG - Fluorodeoxyglucose

GAN - Generative Adversarial Network

GDC - Genomic Data Commons

GDPR - General Data Protection Regulation

GUID - Globally Unique Identifier

HIPAA - Health Insurance Portability and Accountability Act

HL7 - Health Level Seven

IC - Institute or Center

ICC - International Color Consortium

IHC - Immunohistochemistry

IHE - Integrating the Healthcare Enterprise

IOD - Information Object Definition

ISO - International Organization for Standardization

IT - Information Technology

ITI -  IT Infrastructure

JPEG - Joint Photographic Experts Group

JUMBF - JPEG Universal Metadata Box Format

MIDI - Medical Image De-identification

MR - Magnetic Resonance

MRI - Magnetic Resonance Imaging

MRN - Medical Record Number

NIH - National Institutes of Health

NCI - National Cancer Institute

NEMA - National Electrical Manufacturers Association

NIfTI  - Neuroimaging Informatics Technology Initiative

NIST - National Institute of Standards and Technology

NLP - Natural Language Processing

NM - Nuclear Medicine

OCR - Office for Civil Rights

OME - Open Microscopy Environment

PDF - Portable Document Format

PET - Positron Emission Tomography

PHI - Protected Health Information

PRFI - Potentially Reconstructable Facial Information

OCR - Optical Character Recognition

OCT - Optical Coherence Tomography

OID - Object Identifier

OSI - Open Systems Interconnection

PACS - Picture Archiving and Communication System

PRSH - Privacy Rule Safe Harbor

QC - Quality Control

ROI - Region of Interest

RT - Radiotherapy

RTSS - RT Structure Set

SDC - Statistical Disclosure Control

SDTM - Study Data Tabulation Model

SNOMED CT - Systematized Nomenclature of Medicine Clinical Terms

SOP - Service-Object Pair

SPECT - Single Photon Emission Computed Tomography

SR - Structured Report

SSN - Social Security Number

TCGA - The Cancer Genome Atlas

TCIA - The Cancer Imaging Archive

TIFF - Tag(ged) Image File Format

US - Ultrasound

SUV - Standard(ized) Uptake Value

UID - Unique Identifier

UUID - Universally Unique Identifier

VL - Visible Light

VR - Value Representation

WSI - Whole Slide Imaging

XML - Extensible Markup Language

## 1.29 References

All hyperlinks last accessed 2023/03/18.